\documentclass[aps,prc,twocolumn,superscriptaddress,floatfix,amsmath,amssymb,nofootinbib,longbibliography]{revtex4-2}

\usepackage{bm}
\usepackage{dcolumn}
\usepackage{amsfonts}
\usepackage{amsthm}
\usepackage{bbold}
\usepackage{color}
\usepackage{braket}
\usepackage{graphicx}
\usepackage[caption=false]{subfig}
\usepackage{orcidlink}

\usepackage{hyperref}
\hypersetup{
    colorlinks = true,
    citecolor  = blue,
    linkcolor  = blue,
    urlcolor  = blue
}

\newcommand{\magicint}{1.8/2.0 (EM)}
\newcommand{\deltago}{$\Delta$NNLO$_\text{GO}$ (394)}
\newcommand{\emarthuis}{1.8/2.0 (EM7.5)}

\newcommand{\nnlosat}{NNLO$_\text{sat}$}

\newcommand{\ai}{\textit{ab initio}}
\newcommand{\eg}{\textit{e.g.}}
\newcommand{\ie}{\textit{i.e.}}
\newcommand{\Xm}[1]{\ensuremath{#1_\text{max}}}
\newcommand{\MeV}{\ensuremath{\text{MeV}}}

\newcommand{\fm}{\ensuremath{\text{fm}}}

\newcommand{\elem}[2]{\ensuremath{^{#2}\text{#1}}}

\begin{document}

\title{Impact of ground-state correlations on the multipole response of nuclei:\\
Ab initio calculations of moment operators}

\allowdisplaybreaks

\author{A.~Porro \orcidlink{0000-0001-9828-546X}}
\email{andrea.porro@tu-darmstadt.de}
\affiliation{Technische Universit\"at Darmstadt, Department of Physics, 64289 Darmstadt, Germany}
\affiliation{ExtreMe Matter Institute EMMI, GSI Helmholtzzentrum f\"ur Schwerionenforschung GmbH, 64291 Darmstadt, Germany}

\author{A.~Schwenk \orcidlink{0000-0001-8027-4076}}
\email{schwenk@physik.tu-darmstadt.de}
\affiliation{Technische Universit\"at Darmstadt, Department of Physics, 64289 Darmstadt, Germany}
\affiliation{ExtreMe Matter Institute EMMI, GSI Helmholtzzentrum f\"ur Schwerionenforschung GmbH, 64291 Darmstadt, Germany}
\affiliation{Max-Planck-Institut f\"ur Kernphysik, Saupfercheckweg 1, 69117 Heidelberg, Germany}

\author{A.~Tichai \orcidlink{0000-0002-0618-0685}}
\email{alexander.tichai@tu-darmstadt.de} 
\affiliation{Technische Universit\"at Darmstadt, Department of Physics, 64289 Darmstadt, Germany}
\affiliation{ExtreMe Matter Institute EMMI, GSI Helmholtzzentrum f\"ur Schwerionenforschung GmbH, 64291 Darmstadt, Germany}
\affiliation{Max-Planck-Institut f\"ur Kernphysik, Saupfercheckweg 1, 69117 Heidelberg, Germany}

\begin{abstract}
We develop a framework that allows to calculate integrated properties of the nuclear response from first principles. Using the \ai{} in-medium similarity renormalization group (IMSRG), we calculate the expectation values of moment operators that are linked to the multipole response of nuclei.
This approach is applied to the isoscalar mono- and quadrupole as well as the isovector dipole response of closed-shell nuclei from $^4$He to $^{78}$Ni for different chiral two- and three-nucleon interactions.
We find that the inclusion of many-body correlations in the nuclear ground state significantly impacts the multipole response when going from the random-phase approximation to the IMSRG level. 
Our IMSRG calculations lead to an improved description of experimental data in \elem{O}{16} and \elem{Ca}{40}, including a good reproduction of the Thomas-Reiche-Kuhn enhancement factor.
These findings highlight the utility of the moment method as a benchmark for other \ai{} approaches that describe nuclear response functions through the explicit treatment of excited states.
\end{abstract}

\maketitle

\section{Introduction}

Recent years have witnessed remarkable progress in the microscopic description of atomic nuclei from first principles, enabling \ai{} calculations across the nuclear chart and up to \elem{Pb}{208}~\cite{Hergert2020FP_AbInitioReview,Stroberg2021PRL_AbInitioLimits,Hu2021lead}. In addition to this expansion in reach, the scope of nuclear observables accessible through \ai{} methods has grown substantially. These include studies of electromagnetic moments~\cite{Miyagi23a,Acharya:2023ird,King:2024zbv}, weak decays~\cite{Gysb19beta,Belley:2023lec}, and applications to deformed nuclei~\cite{Novario2020,Hagen2022,Duguet22a,Sun24a,Zhou24a}. These advances have been made possible by the synergy between systematically improvable nuclear forces derived from chiral effective field theory (EFT)~\cite{Epelbaum19a,Machleidt24a} and systematic many-body methods capable of treating complex nuclear correlations (see, \eg, Refs.~\cite{Hagen2014rpp,Hergert16a,Tichai2020review,Soma2020SCGF,Hebe203NF}).

Historically, collective excitations in nuclei have been explored using the random-phase approximation (RPA) and the generator coordinate method (GCM), both typically built on phenomenological energy density functionals (EDFs) as well as through Lanczos strength-function techniques within the nuclear shell model~\cite{Ring1980_ManyBodyBook,Bortignon98a,Harakeh01a,Garg18a,Colo22a}. More recently, \ai{} methods have also been employed to investigate nuclear collective responses. These include the Lorentz integral transform method combined with the coupled-cluster ansatz~\cite{Bacca13a,Bacca14a,Miorelli16a,Acharya24a,Bonaiti2024,Marino2025openshell}, the self-consistent Green’s function method~\cite{Raimondi18a,Raimondi18b}, and the equation-of-motion approach built upon the in-medium similarity renormalization group~\cite{Parzuchowski16a,Parzuchowski17a}, all of which proceed via the inclusion of excited states.

Efforts to perform RPA calculations with chiral interactions have been undertaken in symmetry-conserving frameworks~\cite{Papakonstantinou17a,Wu18a,Hu20a} and later generalized to deformed nuclei through the symmetry-breaking quasiparticle RPA~\cite{Beaujeault-Taudiere22a,Zaragoza24a}, highlighting the essential role of static correlations. Moreover, RPA and related approaches built upon correlated ground states have emphasized the importance of consistently treating correlations in both ground and excited states~\cite{Carter22a,Beaujeault-Taudiere22a}. Finally, the projected GCM (PGCM) method, and its extension including perturbative corrections (PGCM-PT), have been developed to simultaneously describe rotational and vibrational collective degrees of freedom~\cite{Frosini21a,Frosini21b,Frosini21c,Porro24a,Porro24b,Porro24c,Porro24d}.

In this work, we develop \ai{} calculations of the multipole response of closed-shell nuclei using the in-medium similarity renormalization group (IMSRG)~\cite{Tsukiyama10a,Hergert15a}. In particular, we investigate moment operators of the isoscalar and isovector multipole response. This allows the extraction of integrated information of the nuclear response from calculations of ground-state expectation values only. This technique is used to study the impact of many-body correlations by comparing IMSRG results to mean-field Hartree-Fock (HF) or RPA calculations.

This work is organized as follows. In Sec.~\ref{sec:theo} we introduce the moment operators and the theoretical methods used, followed by a discussion of sum rules and gauge invariance in Sec.~\ref{sec:sum_rules}. We present results for a broad range of nuclei and multipoles in Sec.~\ref{sec:num_res}, with comparison to experimental data when available. Finally, we summarize our results and future directions in Sec.~\ref{sec:concl}. Details on the derivations and further numerical benchmarks are given in the appendices.

\section{Formalism}
\label{sec:theo}

The evaluation of moments of the nuclear response based on ground-state expectation values has been the subject of extensive works in the past. Excellent reviews cover this wide topic in detail~\cite{Bohigas79a,Lipparini89,Christillin89a,Orlandini91a}. We review the basic ingredients here, in order to show how this method is adapted to the context of \ai{} methods, with a brief focus on the IMSRG used in this work.

\subsection{Moments of the strength function}

In the following, the set $\{\ket{\Psi_\nu}\}_\nu$ denotes the eigenstates of a many-body Hamiltonian $H$ satisfying the time-independent Schr\"odinger equation
\begin{equation}
    H\ket{\Psi_\nu}=E_\nu\ket{\Psi_\nu} \, ,
\end{equation}
where
\begin{equation}
\label{eq:Hamiltonian}
    H=T-T_{\text{cm}}+V_{\text{NN}}+V_{\text{3N}} \, .
\end{equation}
Here, $T$ is the total kinetic energy operator, $T_{\text{cm}}$ the center of mass (cm) kinetic energy, and $V_{\text{NN}}$ denotes nucleon-nucleon (NN) and $V_{\text{3N}}$ three-nucleon (3N) interactions. Let $\ket{\Psi_0}$ refer to the many-body ground state, with $E_0$ the corresponding ground-state energy. 

Given a set of operators $Q_{\lambda\mu}$, which behave under rotation like the $\mu$-th component of a spherical tensor~\cite{Varshalovich88a,Rose95a,Suhonen07a,Blatt52a,DeShalit13a} of rank $\lambda$, the associated strength function reads as
\begin{equation}
    S(Q_{\lambda},E)\equiv\sum_{\mu\nu}|\braket{\Psi_\nu|Q_{\lambda\mu}|\Psi_0}|^2 \, \delta(E_\nu-E_0-E) \,.\label{eq:strength}
\end{equation}
It describes the transition probability between the ground and excited states and fully characterizes the linear response of the nucleus to a perturbation induced by $Q_\lambda$. Using the Wigner-Eckart theorem, Eq.~\eqref{eq:strength} may also be rewritten in terms of reduced matrix elements
\begin{equation}
    S(Q_{\lambda},E)=\frac{1}{2J_0+1}\sum_{\nu}|\braket{\Psi_\nu||Q_{\lambda}||\Psi_0}|^2 \, \delta(E_\nu-E_0-E) ,
\end{equation}
where $J_0$ is the total angular momentum of $\ket{\Psi_0}$.

The moments of the strength function are then defined
\begin{align}
    m_k(Q_\lambda)&\equiv\int E^k S(Q_\lambda, E)dE\nonumber\\
    &=\sum_{\mu\nu}(E_\nu-E_0)^k|\braket{\Psi_\nu|Q_{\lambda\mu}|\Psi_0}|^2\,.\label{eq:mom_def}
\end{align}
They provide compact information about the strength function and correspond to the mathematical definition of the $k$-th moment of a discretized probability distribution. Different moments $m_k(Q_\lambda)$ of the strength function $S(Q_\lambda,E)$ are related through inequalities~\cite{Bohigas79a}, whose validity is founded on the positivity of $S(Q_\lambda,E)$. In particular, it is possible to show, using Schwartz's inequality, that
\begin{equation}\label{eq:ineq_chain}
    \ldots\leq\sqrt{\frac{m_{k+1}}{m_{k-1}}}\leq\frac{m_{k+1}}{m_{k}}\leq\sqrt{\frac{m_{k+2}}{m_{k}}}\leq\frac{m_{k+2}}{m_{k+1}}\leq\ldots\,,
\end{equation}
where the $Q_{\lambda}$-label has been dropped for the sake of brevity. All these ratios carry units of energy, which suggests the definition of a set of average energies as
\begin{subequations}
\label{eq:averages}
    \begin{align}
        \tilde{E}_k(Q)&\equiv\sqrt{\frac{m_k(Q)}{m_{k-2}(Q)}}\,,\\
        \bar{E}_k(Q)&\equiv\frac{m_k(Q)}{m_{k-1}(Q)}\,.
    \end{align}
\end{subequations}
The equalities in Eq.~\eqref{eq:ineq_chain} would hold if the strength were entirely concentrated in a single peak. The spread of the different averages in Eqs.~\eqref{eq:averages} reflects quantitatively the fragmentation of the strength $S(Q_\lambda,E)$.

\subsection{Ground-state method for moments}

Equation~\eqref{eq:mom_def} relates the moments of the strength function to the knowledge of the entire spectrum of $H$, such that the evaluation of the moments in principle entails the full diagonalization of $H$. Modern diagonalization solvers are based on the Lanczos method, an iterative algorithm used to diagonalize large, sparse Hermitian matrices. By constructing a Krylov subspace from an initial vector, the Lanczos method reduces the Hamiltonian to a tridiagonal form whose eigenvalues approximate the extremal eigenvalues of the original matrix. Importantly, the Lanczos algorithm preserves the moments of the strength function, such that the first $k$ iterations of the Lanczos method reproduce the lowest $2k$ moments of the strength function. However, diagonalization techniques scale poorly with respect to the mass number $A$, so other strategies must be explored.

An alternative strategy successfully used in previous investigations is based on ground-state expectation values only~\cite{Johnson15a,LuJohnson2018,Johnson20a,Porro24c,Xie25a}. In this framework, the knowledge of all the excited states $\ket{\Psi_\nu}$, which are necessary for the calculation of Eq.~\eqref{eq:mom_def}, is replaced by the evaluation of the expectation value of a (complicated) operator over the ground state $\ket{\Psi_0}$. Indeed, by invoking the closure relation
\begin{equation}
\label{eq:closure}
    \mathbb{1}=\sum_\nu\ket{\Psi_\nu}\bra{\Psi_\nu}\,,
\end{equation}
Eq.~\eqref{eq:mom_def} can be rewritten as
\begin{align}
    m_k(Q_\lambda)&=\sum_\mu\braket{\Psi_0|Q^\dagger_{\lambda\mu}[\underbrace{{H},...[{H},[{H}}_{\text{$k$ times}},{Q_{\lambda\mu}}]]...]|\Psi_0}\,,
\end{align}
where the ground-state property
\begin{equation}
    (H-E_0)^k\ket{\Psi_0}=0
\end{equation}
has been exploited in order to introduce the nested commutators. Moreover, for the $\mu$-th component of a spherical tensor of rank $\lambda$, the following relation holds
\begin{equation}
    Q_{\lambda\mu}^\dagger=(-1)^\mu Q_{\lambda-\mu}\,,
\end{equation}
such that the odd-$k$ moments can be further simplified as
\begin{multline}
        m_k(Q_\lambda)= \\ \frac{1}{2}\sum_\mu(-1)^\mu \braket{\Psi_0|[Q_{\lambda-\mu},[\underbrace{{H},...[{H},[{H}}_{\text{$k$ times}},{Q_{\lambda\mu}}]]...]]]|\Psi_0}\,.
\end{multline}
This provides a very useful simplification: If $H$ is a two-body Hamiltonian, then the odd-$k$ moments $m_k(Q_\lambda)$ are given by the expectation value of a $(k+1)$-body operator, while the even ones, for which the simplification does not apply, are the expectation value of a $(k+2)$-body operator.

In this work, we thus focus on the zeroth and first moments, which can both be expressed in terms of the expectation value of two-body operators
\begin{subequations}
\label{eq:mom_op_gen}
    \begin{align}
        m_0(Q_\lambda)&=\braket{\Psi_0|M_0(Q_\lambda)|\Psi_0}\,,\\
        m_1(Q_\lambda)&=\braket{\Psi_0|M_1(Q_\lambda)|\Psi_0}\,,
    \end{align}
\end{subequations}
where the zeroth and first moment operators are given by\footnote{Equations~\eqref{eq:mom_op_gen} hold true for transition operators with vanishing expectation value in the ground state. If this assumption is not true, the expectation value must be subtracted~\cite{Lipparini89}. This is the case for the monopole operator $r^2$, where
\begin{equation}
    m_0(r^2)=\braket{\Psi_0|r^2\cdot r^2|\Psi_0}-|\braket{\Psi_0|r^2|\Psi_0}|^2\,. \nonumber
\end{equation}}
\begin{subequations}
\label{eq:moms_op}
    \begin{align}
        M_0(Q_\lambda)&\equiv\sum_\mu(-1)^\mu Q_{\lambda-\mu}Q_{\lambda\mu}\,,\label{eq:m0_op}\\
        M_1(Q_\lambda)&=\frac{1}{2}\sum_\mu(-1)^\mu[Q_{\lambda-\mu},[H,Q_{\lambda\mu}]]\label{eq:double_comm}\,.
    \end{align}
\end{subequations}
Despite involving operators that behave like spherical tensors under rotation, the operators in Eq.~\eqref{eq:moms_op} are coupled to give scalar operators. With the expressions of the matrix elements from Ref.~\cite{LuJohnson2018}, the operators entering Eqs.~\eqref{eq:moms_op} have been implemented within the \textsc{IMSRG++} code~\cite{Stroberg2024_IMSRGGit}. The $m_0$ and $m_1$ moments being the only quantities that can be expressed as the expectation value of two-body operators (if an effective two-body Hamiltonian $H$ and one-body operator $Q_\lambda$ are considered), the present study targets the systematic evaluation of such quantities across the nuclear chart using the IMSRG method. It is important to stress that, because operators in Eq.~\eqref{eq:moms_op} derive from the exact closure relation from Eq.~\eqref{eq:closure}, they fully span the Hilbert space, with the only approximation being given by its finite-basis representation. The used ground state, on the other hand, is the outcome of an approximate many-body calculation. 

\subsection{The IMSRG method}

The idea of the IMSRG~\cite{Tsukiyama10a,Hergert15a} is to evolve the initial Hamiltonian normal-ordered to a reference state, $H(0)=H$, via a continuous series of unitary transformations $U(s)$ to decouple particle-hole excitations on top of the reference state,
\begin{equation}
    H(s)=U(s)H(0)U^\dagger(s)\, .
\end{equation}
The series of unitary transformations can be cast as a flow equation with the flow parameter $s$
\begin{equation}
    \frac{dH(s)}{ds}=[\eta(s),H(s)]\,,
\end{equation}
where $\eta(s)$ is the anti-Hermitian generator of the transformation
\begin{equation}
    \eta(s)=\frac{d\,U(s)}{ds}U^\dagger(s)=-\eta^\dagger(s)\,.
\end{equation}
In this work the generator is chosen as the imaginary-time generator that is widely used in applications to closed-shell nuclei.

The key of the IMSRG is to start from a many-body reference state $\ket{\Phi_0}$, which we take to be the Hartree-Fock (HF) ground state. All operators are then normal ordered with respect to $\ket{\Phi_0}$, such that the zero-body component of the Hamiltonian after normal ordering is given by its expectation value $\braket{\Phi_0|H(0)|\Phi_0}$. The flow suppresses the coupling of $\ket{\Phi_0}$ to particle-hole excitations, such that the zero-body term approaches the exact ground-state energy in the large $s$ limit (if all operators in the IMSRG flow are kept),
\begin{equation}
    \lim_{s\to\infty} \braket{\Phi_0|H(s)|\Phi_0}=E_0\,.
\end{equation}
In this work, the Magnus formulation of the unitary transformation is used~\cite{Morris15a}. The Magnus expansion provides a solution of the form
\begin{equation}
    U(s)=e^{\Omega(s)} \,,
\end{equation}
which has the advantage of enforcing the unitarity of the transformation at all stages of the flow even in approximated calculations. 

With this, the transformed Hamiltonian, as well as any other operator $O$, can be constructed as
\begin{subequations}
    \begin{align}
        H(s)&=e^{\Omega(s)}He^{-\Omega(s)}\,,\\
        O(s)&=e^{\Omega(s)}Oe^{-\Omega(s)}\,.
    \end{align}
\end{subequations}
This also extends to the moment operators defined in Eqs.~\eqref{eq:moms_op}, such that the moments expectation value at the HF and IMSRG level are given by
\begin{subequations}
    \label{eq:used_mom_def}
    \begin{align}
        m_k(Q_\lambda,0)&=\braket{\Phi_0|M_k(Q_\lambda)|\Phi_0}\,,\\
        m_k(Q_\lambda,s)&=\braket{\Phi_0|e^{\Omega(s)}M_k(Q_\lambda)e^{-\Omega(s)}|\Phi_0}\,,
    \end{align}    
\end{subequations}
respectively. Thus, differences in the value of $m_k$ before (HF) and after (IMSRG) the flow are indicative of the impact of ground-state correlations on the nuclear response.

In this work, we truncate all commutators at the normal-ordered two-body level giving rise to the IMSRG(2) truncation.
While three-body operators can be included albeit at substantial numerical cost, the IMSRG(2) truncation provides very reliable results for ground-state properties of medium-mass nuclei~\cite{Heinz:2024juw}.

\section{Sum rules and gauge invariance}
\label{sec:sum_rules}

\subsection{Sum rules}

The evaluation of moments of the strength distribution from ground-state expectation values of operators of the form~\eqref{eq:moms_op} has been addressed in a few cases within the phenomenological shell model (see Refs.~\cite{Johnson15a,LuJohnson2018,Johnson20a,Xie25a}) and the PGCM~\cite{Porro24c}. A technique that has been used more extensively in the past is the so-called sum rule method, especially concerning the first moment of the strength function, the energy-weighted sum rule (EWSR). Within the sum rule method~\cite{Bohigas79a,Lipparini89}, often one assumes that nuclear interactions are local, \ie{}, $V=V(\vec{r}_1-\vec{r}_2)$. In this case, the double commutator structure in Eq.~\eqref{eq:double_comm} simplifies to
\begin{align}
\label{eq:edf_assumption}
    \text{EWSR}(Q_\lambda(\vec{r}))&=\frac{1}{2}\braket{\Psi_0|[Q_{\lambda}^\dagger(\vec{r}),[T+V,Q_{\lambda}(\vec{r})]]|\Psi_0}\nonumber\\
    &=\frac{1}{2}\braket{\Psi_0|[Q_{\lambda}^\dagger(\vec{r}),[T,Q_{\lambda}(\vec{r})]]|\Psi_0}\,.
\end{align}
In the following, we use the expression ``sum rules'' to indicate that the EWSR has been evaluated assuming that Eq.~\eqref{eq:edf_assumption} applies. In this case, the double commutator depends only on the kinetic energy, and EWSRs lead to simple relations between the first moment of isoscalar operators and properties of the nuclear ground state. We note that this is not the case for the nuclear Hamiltonians we consider. Traditionally, EWSRs have been derived in the laboratory frame, which is the most frequent choice in EDF applications. In this work, on the other hand, the intrinsic Hamiltonian~\eqref{eq:Hamiltonian} is employed and the sum rules must be adapted consequently. This yields different definitions for the EWSRs of isoscalar multipole operators (see App.~\ref{sec:app_intrinsic} for details). The final expressions for the isoscalar monopole and quadrupole intrinsic EWSRs are
\begin{subequations}
\label{eq:srs}
    \begin{align}
        \text{EWSR}_{\,\text{int}}(r^2)&=\frac{2\hbar^2}{m}\braket{\Psi_0|r^2_\text{int}|\Psi_0}\,,\label{eq:sr_0}\\
        \text{EWSR}_{\,\text{int}}(Q_2)&=\frac{25}{4\pi}\frac{\hbar^2}{m}\braket{\Psi_0|r^2_\text{int}|\Psi_0}\,,\label{eq:sr_2}
    \end{align}
\end{subequations}
with $r^2_\text{int}$ the intrinsic mean-square radius operator defined in Eq.~\eqref{eq:r_int} and $m$ the nucleon mass. More general definitions for higher multipoles can be found in App.~\ref{sec:app_intrinsic}. 

The EWSR of the isovector dipole operator has also been widely studied. The dipole mode is by far the most dominant contribution to the electromagnetic excitation of nuclear systems and the associated sum rule is directly proportional to the total photoabsorption cross section. When dealing with electric excitations, however, the same simplifications leading to the omission of the interaction contribution in the isoscalar EWSR do not apply, due to the presence of meson-exchange currents or tensor forces. If these are neglected, an expression analogous to the Thomas-Reiche-Kuhn (TRK) sum rule~\cite{Thomas25a,Kuhn25a} in atomic systems can be obtained (see also App.~\ref{sec:app_intrinsic})
\begin{equation}
    \text{EWSR}_{\text{int}}(E1)=\frac{\hbar^2e^2}{2m}\frac{9}{4\pi}\frac{NZ}{A}\,,
    \label{eq:trk}
\end{equation}
where $e$ is the electric charge, $N$ and $Z$ denote neutron and proton numbers, respectively, and $A$ is the total mass number.

\subsection{Gauge invariance and continuity equation}

In the EDF framework, locality is equivalent to invariance under local gauge transformations~\cite{Blaizot86a}, an essential condition for the system to couple consistently to electromagnetic fields. The potential breaking of gauge invariance has been extensively studied by the EDF community~\cite{Dobaczewski95a,Dobaczewski95a-erratum,Carlsson08a,Carlsson08a-erratum,Raimondi11a,Hinohara19a}. In Ref.~\cite{Porro24c}, a study using chiral EFT interactions was performed in the spirit of Ref.~\cite{Hinohara19a}, but a detailed investigation is still missing. In this section, a brief overview of the implications of gauge invariance is presented, following the discussion of Ref.~\cite{Arenhovel93a}.
 
It can be shown~\cite{Arenhovel80a,Arenhovel93a} that local $U(1)$ gauge invariance
leads to the continuity equation for the electromagnetic current and charge
\begin{equation}
    \vec{\nabla}\cdot\vec{j}(\vec{r},t)+\frac{i}{\hbar}[H,\rho(\vec{r},t)]+\partial_t\rho(\vec{r},t)=0\,,
    \label{eq:continuity}
\end{equation}
where $\rho$ and $\vec{j}$ are the charge density and current, respectively. The same Eq.~\eqref{eq:continuity} also holds if $\rho$ and $\vec{j}$ are assumed to be the probability density and current, in which case it implies the probability conservation, extending the present discussion to isoscalar operators as well. For time-independent densities and currents one finds
\begin{equation}
\label{eq:tind}
    \vec{\nabla}\cdot\vec{j}(\vec{r})+\frac{i}{\hbar}[H,\rho(\vec{r})]=0\,.
\end{equation}
Equation~\eqref{eq:tind} can be decomposed according to its many-body rank, defining a hierarchy of continuity equations
\begin{equation}
    \vec{\nabla}\cdot\vec{j}_{[n]}(\vec{r})+\frac{i}{\hbar}\sum_{k=1}^n[H_{[n+1-k]},\rho_{[k]}(\vec{r})]_{[n]}=0\,,
    \label{eq:nbody}
\end{equation}
where the subscript $\cdot_{[n]}$ indicates an $n$-body operator. The one-body part of Eq.~\eqref{eq:nbody} is given by 
\begin{equation}
    \vec{\nabla}\cdot\vec{j}_{[1]}(\vec{r})+\frac{i}{\hbar}[T,\rho_{[1]}(\vec{r})]=0\,,
    \label{eq:cont_1}
\end{equation}
which holds for the usual definitions of the one-body charge and current densities~\cite{Blaizot86a}. Equation~\eqref{eq:cont_1} is then the gauge invariance condition for a local potential.

However, for interactions involving the exchange of  particles (mesons) that also couple to the electromagnetic field or for nonlocal interactions, one needs in addition higher many-body currents for the gauge condition to be satisfied (see, \eg, Refs.~\cite{Christillin89a,Walecka01a}), as visible from Eq.~\eqref{eq:nbody}. Assuming the charge density does not have many-body contributions~\cite{Christillin89a}, \ie{}, $\rho_{[n]}(\vec{r}) = 0$ for $n\geq2$ (Siegert's hypothesis~\cite{Siegert37a,Arenhovel93a}), then from Eq.~\eqref{eq:nbody} one only needs additional many-body contributions to $\vec{j}(\vec{r})$, implying
\begin{equation}
    \vec{\nabla}\cdot\vec{j}_{[n]}(\vec{r})+\frac{i}{\hbar}[V_{[n]},\rho_{[1]}(\vec{r})]=0\,.
\end{equation}
The present discussion is relevant in this context because the $m_1$ moment can also be expressed in terms of the expectation value of a similarity-transformed Hamiltonian
\begin{align}
    m_1(Q_\lambda)&=\frac{1}{2}\frac{\partial^2}{\partial\eta^2}\sum_\mu\braket{\Psi_0|e^{-i\eta Q_{\lambda\mu}^\dagger}H\,e^{i\eta Q_{\lambda\mu}}|\Psi_0}\bigg|_{\eta=0}\nonumber\\
    &=\frac{1}{2}\frac{\partial^2}{\partial\eta^2}\sum_\mu\braket{\Psi_{\lambda\mu}(\eta)|H|\Psi_{\lambda\mu}(\eta)}\bigg|_{\eta=0}\,,
\end{align}
where in the second equality the definition
\begin{equation}
    \ket{\Psi_{\lambda\mu}(\eta)}\equiv e^{i\eta Q_{\lambda\mu}}\ket{\Psi_0}
\end{equation}
has been introduced. This shows that $m_1(Q_\lambda)$ is related to the expectation value of $H$ in $\ket{\Psi_{\lambda\mu}(\eta)}$, which can be viewed as the action of a local gauge transformation on the nuclear ground state. Therefore, when evaluating $m_1(Q_{\lambda})$, deviations from the EWSR$(Q_{\lambda})$ evaluated via Eq.~\eqref{eq:edf_assumption}, are not only due to nonlocal interactions but also due to two- and higher-body currents, that are absent in the "classical" picture but are required by the continuity equation~\eqref{eq:tind}.

Finally, we note that in chiral EFT, interactions and currents are derived consistently from the chiral Lagrangian (see, \eg, Ref.~\cite{Krebs20a}). While it has been shown that the leading two-body currents can be obtained from the commutator expression involved in $m_1$~\cite{Riska:1972zz,Christillin89a}, this may be more challenging for consistent higher-order currents.

\section{Numerical results}
\label{sec:num_res}

In this work we study the $m_0$ and $m_1$ moments of the response function, as defined in Eqs.~\eqref{eq:used_mom_def}, in closed-shell nuclei from $^4$He to $^{78}$Ni. The operators under investigation are the isoscalar (IS) and isovector (IV) multipole operators $Q^{(\text{IS/IV})}_{\lambda \mu}$:
\begin{subequations}
    \begin{align}
        Q^\text{IS}_{00}&=\sum_{i=1}^Ar_i^2 \, ,  \\
        Q^\text{IV}_{1\mu}&=\frac{N}{A}\sum_{i=1}^Zr_iY_{1\mu}(\hat{r}_i)-\frac{Z}{A}\sum_{i=1}^Nr_iY_{1\mu}(\hat{r}_i) \, ,\label{eq:dipole_int}\\
        Q^\text{IS}_{2\mu}&=\sum_{i=1}^Ar_i^2Y_{2\mu}(\hat{r}_i) \, ,
    \end{align}
\end{subequations}
where $Y_{\lambda\mu}$ denote spherical harmonics and the coordinates $\vec{r}_i$ are considered in the laboratory frame. Expressions for the matrix elements of the above operators in a spherical harmonic-oscillator (HO) basis can be found, \eg, in Refs.~\cite{DeForest66a,Donnelly80a,Ring1980_ManyBodyBook,Suhonen07a}, while a discussion about the use of laboratory or intrinsic coordinates is found in App.~\ref{sec:app_intrinsic}.

We present results at the HF and IMSRG(2) levels. In some cases, we also include RPA results, given the important role of the RPA framework for the description of collective nuclear excitations. All results have been obtained using the \textsc{IMSRG++} code~\cite{Stroberg2024_IMSRGGit}. We note that, because of the Thouless theorem~\cite{Thouless61a,Bohigas79a,Ring1980_ManyBodyBook,Capelli09a}, the $m_1$ moment, as evaluated using Eq.~\eqref{eq:double_comm} for the HF ground state, is equivalent to a full RPA calculation. Thus, as far as the $m_1$ moment is concerned, RPA and HF are equivalent. The same statement does not hold for the $m_0$ moment, where the expectation value of the operator~\eqref{eq:m0_op} at the HF level and in Tamm-Dancoff approximation (TDA) are equivalent~\cite{Ring1980_ManyBodyBook,Suhonen07a}. We have numerically checked the HF-RPA equivalence for $m_1$ and the TDA-HF one for $m_0$ as a benchmark to ensure the correct implementation of the operators~\eqref{eq:moms_op}.

\begin{figure*}[t!]
    \centering
    \includegraphics[width=\textwidth]{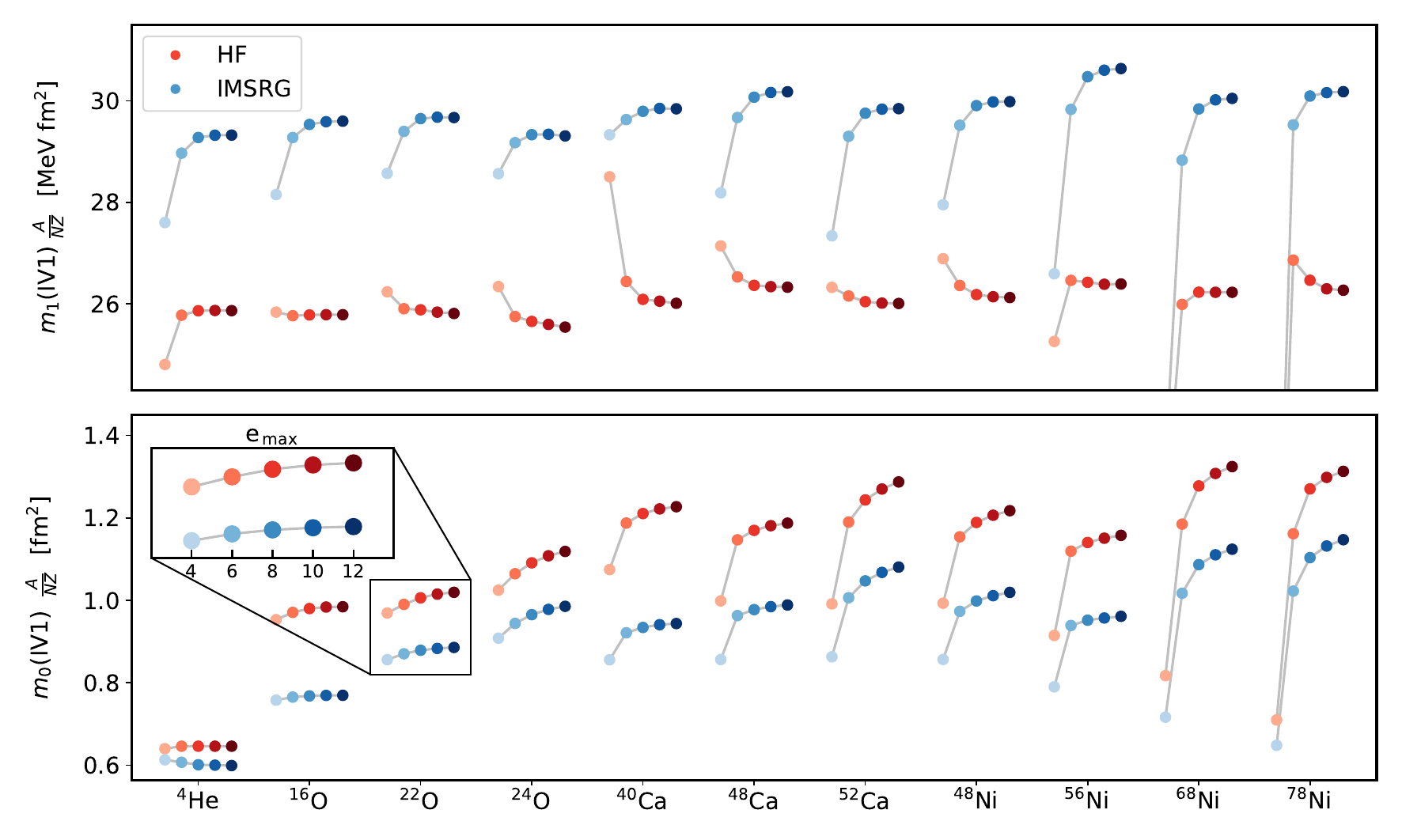}
    \caption{Convergence of $m_1$ (top panel) and $m_0$ (bottom panel) for the isovector dipole (IV1) response as a function of the model space size $\Xm{e}$ at the HF and IMSRG(2) level. For each isotope, the results with increasing $\Xm{e}$ are shown, as highlighted in the inset. The values of $m_1$ and $m_0$ are rescaled by a factor $\tfrac{A}{NZ}$ in order to show all nuclei on the same scale [see Eq.~\eqref{eq:trk}]. Results are given for the 1.8/2.0 (EM) interaction using $\hbar\omega=16$\,MeV.}
    \label{fig:L1_emax_conv}
\end{figure*}

\begin{figure*}[t!]
    \centering
    \includegraphics[width=\textwidth]{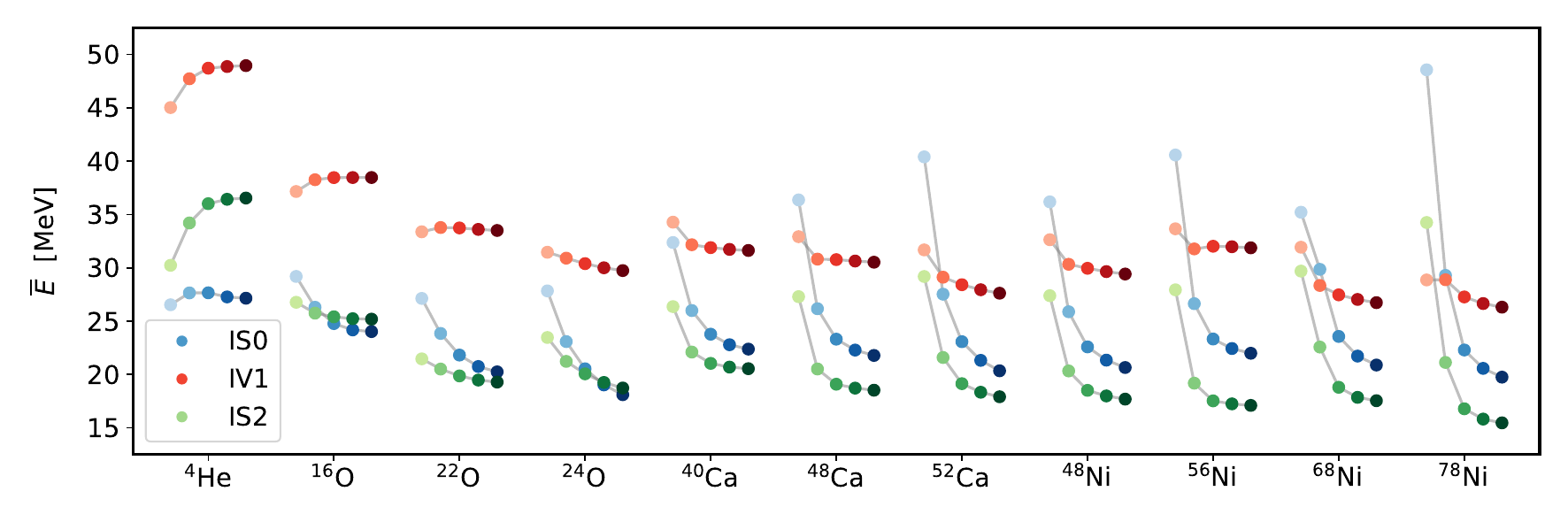}
    \caption{Convergence of the average energy $\bar{E}=m_1/m_0$ as a function of the model space size $\Xm{e}$ at the IMSRG(2) level for different multipolarities. Results are given for the 1.8/2.0 (EM) interaction using $\hbar\omega=16$\,MeV.}
    \label{fig:E_e_conv}
\end{figure*}

\subsection{Computational setup}
\label{sec:setup}

The moments of the nuclear response are studied for a set of chiral interactions. In this work we employ the \magicint{} interaction from Ref.~\cite{Hebeler10a}, the \nnlosat{} interaction from Ref.~\cite{Ekstrom15a,Ekstrom15a-erratum}, the \deltago{} interaction from Ref.~\cite{Jiang20a} as well as the recently developed \emarthuis{} interaction from Ref.~\cite{Arthuis24a}.
The NN and 3N interactions are expanded in spherical HO basis states consisting of 13 major shells, \ie{}, $\Xm{e} = (2n + l)_\text{max} = 12$. Three-nucleon interactions are included using the normal-ordered two-body approximation to avoid the handling of explicit three-body operators during the IMSRG flow~\cite{Hergert15a,Heinz2020,Frosini21d}. This has been shown to introduce only small errors in medium-mass nuclei~\cite{Roth12,Heinz:2024juw}.
To ensure convergence in heavy systems the number of three-body configurations is allowed up to $e_1 + e_2 + e _3 \leq E^{(3)}_\text{max}=24$ by exploiting the developments of Ref.~\cite{Miyagi2022PRC_NO2B}. The underlying HO frequency is taken in a range $\hbar \omega = 12-20 \, \MeV$ depending on the HF-variational minimum for a given nucleus. The NN and 3N matrix elements are computed using the \textsc{NuHamil} code~\cite{Miyagi2023EPJA_NuHamil}.

\begin{figure*}[t!]
    \centering
    \includegraphics[width=\textwidth]{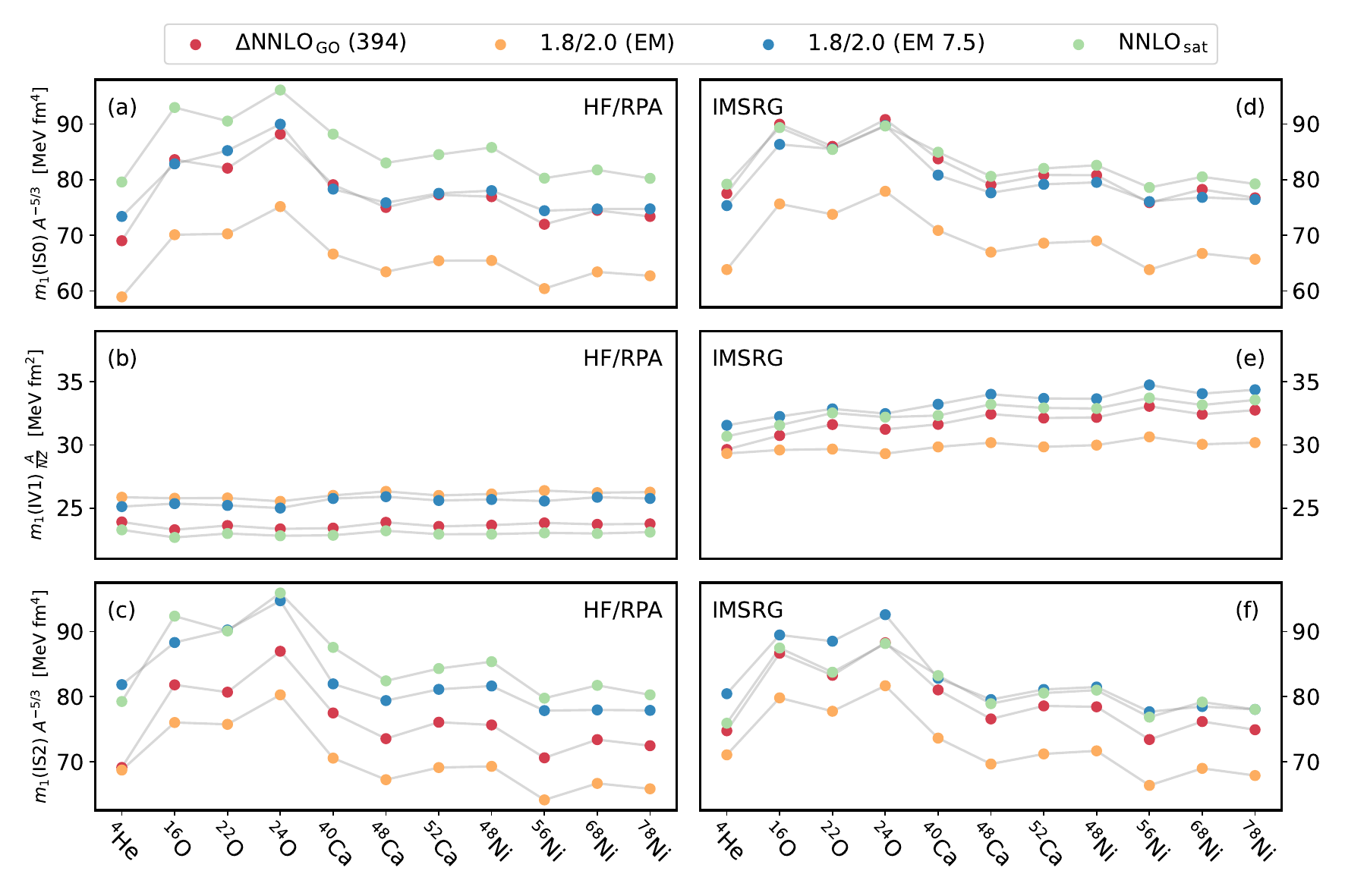}
    \caption{Values of $m_1$ for the isoscalar monopole (IS0, top panels), isovector dipole (IV1, middle panels), and isoscalar quadrupole (IS2,  bottom panels) response for HF/RPA (left) and IMSRG(2) (right) calculations. Results are given for four different interactions (\deltago{}~\cite{Jiang20a}, \magicint{}~\cite{Hebeler10a}, \emarthuis{}~\cite{Arthuis24a}, and \nnlosat{}~\cite{Ekstrom15a}) for $\Xm{e}=12$ and $\hbar\omega=$16\,MeV.}
    \label{fig:int_m1}
\end{figure*}

\subsection{Model-space convergence}
\label{sec:ms_conv}

We first validate that our results are robust in terms of model-space size. To this end, we vary the size of the single-particle basis from $\Xm{e}=4,...,12$ and study the impact on the $m_1$ and $m_0$ moments for the IV dipole response. Results are shown for the \magicint{} interaction in Fig.~\ref{fig:L1_emax_conv}, rescaled by a factor $A/(NZ)$ [see Eq.~\eqref{eq:trk}] to remove the trivial dependence on proton and neutron number. For all nuclei a rapid convergence of $m_1$ as a function of $\Xm{e}$ is observed (top panel) and a model space with $\Xm{e}=12$ is sufficient to obtain converged results for all nuclei, with variations smaller than 0.2\% going from $\Xm{e}=10$ to $\Xm{e}=12$, both at the HF and IMSRG(2) level, even for the heaviest isotopes studied. The convergence of $m_0$ (bottom panel) is, on the other hand, slower. Even if the relative difference between $\Xm{e}=10$ and $\Xm{e}=12$ results is smaller than 1.3\%, one can observe that heavier and neutron rich nuclei do not display a fully converged trend, as in, \eg, $^{24}$O and $^{52}$Ca.
This finding seems to be independent of the many-body truncation and holds equally for the HF and IMSRG calculations. The faster convergence for $m_1$ is likely because operators that can be expressed in terms of commutators benefit from error cancellations~\cite{Rowe2010a}. On the other hand, the slower convergence of $m_0$ shows the dependence of such quantities on the ground-state structure. In $^{24}$O and $^{52}$Ca it could be a signal of continuum effects in the collective response of neutron-rich systems, see, \eg, Ref.~\cite{Hu19a}.
Similar findings hold for the $m_0$ and $m_1$ moments of other operators, such as the isoscalar monopole (IS0) and isoscalar quadrupole (IS2), whose results are reported and briefly discussed in App.~\ref{app:ms_conv}.

As shown in Fig.~\ref{fig:E_e_conv}, the evaluation of the centroid energy $\overline E = m_1/m_0$, instead, is more robust in heavier systems, such that the variations in the value of $m_0$ less strongly affect the ratio itself. We emphasize that the dipole centroid displays a faster convergence with respect to \Xm{e}, the largest error being 1.3\% in $^{52}$Ca, than isoscalar probes (6\% and 4\% error for IS0 and IS2 in $^{24}$O). We attribute this difference to the differential nature of the isovector dipole excitation itself, in which protons and neutrons interfere destructively within the giant dipole resonance.

In addition, we have checked the dependence of the results on the choice of the HO frequency $\hbar\omega$, for details see App.~\ref{app:ms_conv}. For completeness, we report here the relative error (\ie{}, the relative standard deviation) between the results of calculations performed at $\hbar\omega$=12, 16, and 20\,MeV. In the isovector dipole channel the largest deviation is in $^{78}$Ni, with a value of 0.23\% and 0.17\% at the HF and IMSRG level, respectively, while in the same nucleus the $m_0$ error is about 2\%. 

Overall, in the isovector dipole channel we estimate for $m_0$ an error of 1\% and 2\% due to \Xm{e} truncations and $\hbar\omega$ dependence, respectively. For $m_1$ the error is smaller than 0.2\% in both cases. The two errors being correlated we assess a comprehensive 2\% error for $m_0$ and 0.3\% for $m_1$ from the finite basis space employed in the calculations.

\subsection{Interaction dependence and correlation effects}

In the following, all many-body calculations are performed in a $\Xm{e}=12$ model space using a frequency of $\hbar \omega=16\, \MeV$. We discuss the $m_1$ moments for the IS0, IV1, and IS2 operators using the four different chiral interactions introduced in Sec.~\ref{sec:setup}. Results are displayed in Fig.~\ref{fig:int_m1} rescaled by the proton and neutron number dependence of the corresponding sum rules (see Eqs.~\eqref{eq:srs} and~\eqref{eq:trk}), both for HF and IMSRG(2) calculations. 

We observe similar trends for all multipolarities: the \magicint{} interaction gives the lowest values for the IS0 and IS2 operators both at the HF/RPA and IMSRG level, due to the smaller nuclear radius for this interaction. The other three interactions typically yield significantly higher values, with \nnlosat{} giving the largest results for most nuclei (except for light nuclei using the IMSRG). Interestingly, excluding the \magicint{}, the spread of the results from different interactions is strongly reduced passing from HF to IMSRG(2) calculations, with an average relative spread (standard deviation) dropping from 6\% to 2\% in all multipolarities. This effect is consistent with the more converged nature of IMSRG calculations compared to HF/RPA calculations and, importantly, it reduces the interaction dependence of the results. Since integrated cross sections of scattering experiments are an observable proportional to $m_1$, a small interaction dependence is crucial in solid theoretical predictions of such quantities. With the exception of the \magicint{} interaction, which is known to poorly reproduce nuclear radii, \deltago{}, \nnlosat{}, and \emarthuis{} produce consistent results.

\begin{figure}[t!]
    \centering
    \includegraphics[width=\columnwidth]{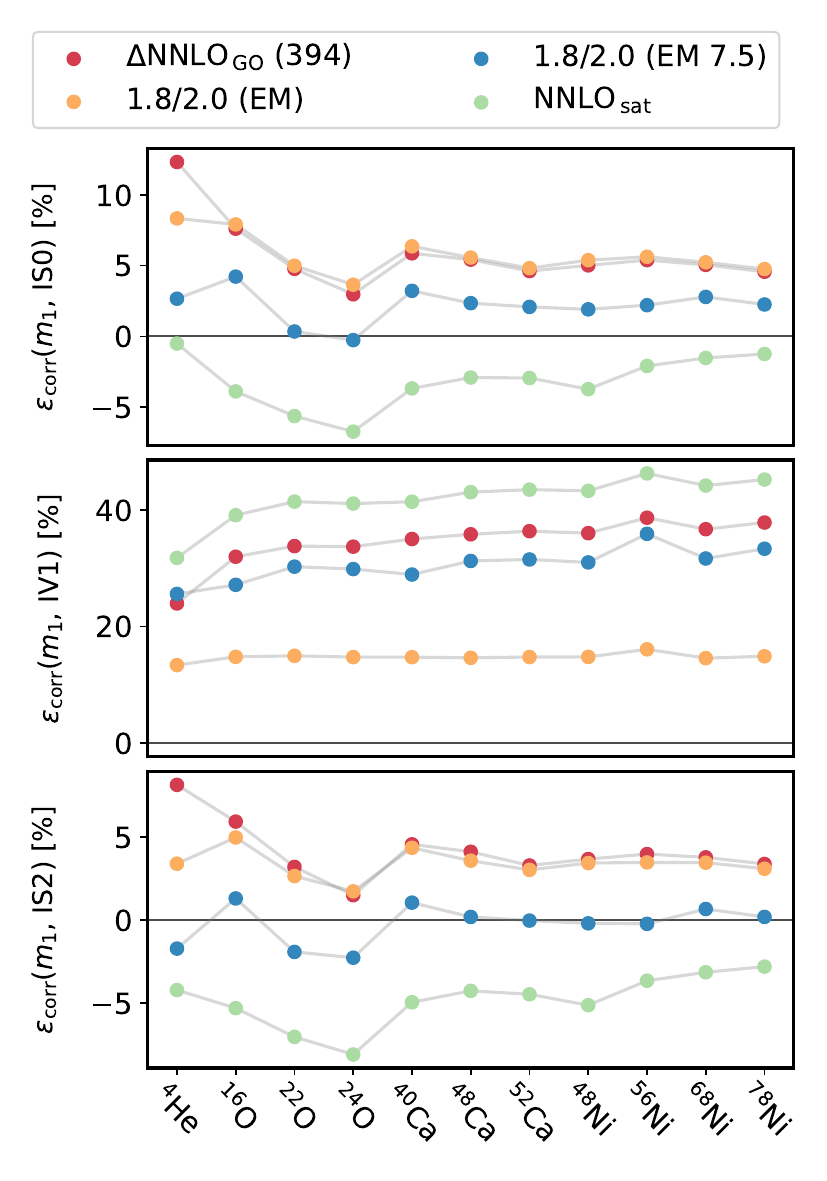}
    \caption{Relative difference between RPA and IMSRG(2) results, as defined by Eq.~\eqref{eq:eps_corr}, for $m_1$ in the IS0 (top panel), IV1 (middle panel), and IS2 (bottom panel) channels. Results are compared for four different interactions (\deltago{}~\cite{Jiang20a}, \magicint{}~\cite{Hebeler10a}, \emarthuis{}~\cite{Arthuis24a}, and \nnlosat{}~\cite{Ekstrom15a}) for $\Xm{e}=12$ and $\hbar\omega=$16\,MeV.}
    \label{fig:int_corr_m1}
\end{figure}

\begin{figure*}[t!]
    \centering
    \includegraphics[width=\textwidth]{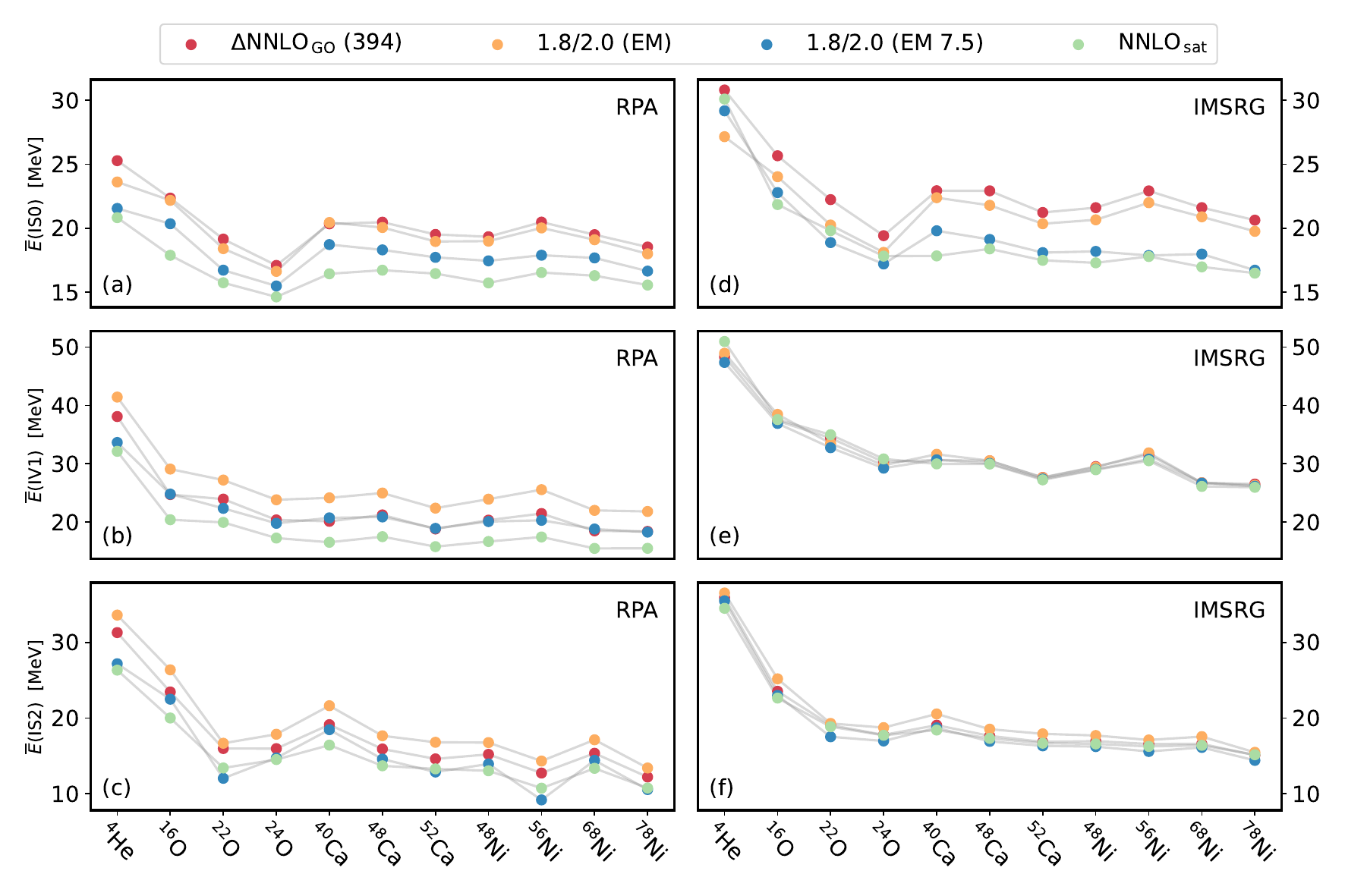}
    \caption{Ratio $m_1/m_0$ at the RPA (left panels) and IMSRG(2) level (right panel) in the IS0 (top), IV1 (middle), and IS2 (bottom) channels. Results are given for four different interactions (\deltago{}~\cite{Jiang20a}, \magicint{}~\cite{Hebeler10a}, \emarthuis{}~\cite{Arthuis24a}, and \nnlosat{}~\cite{Ekstrom15a}) for $\Xm{e}=12$ and $\hbar\omega=$16\,MeV}
    \label{fig:int_E}
\end{figure*}

The impact of many-body correlations is stronger for the IV1 response than for the IS0 and IS2 case, as can be seen in Fig.~\ref{fig:int_m1} comparing the left and right panels. A more quantitative analysis necessitates to isolate the contribution of many-body correlations. This is achieved by evaluating the relative difference
\begin{equation}
    \varepsilon_{\text{corr}}[\%]\equiv\frac{x_{\text{IMSRG}}-x_{\text{RPA}}}{x_{\text{RPA}}}\times 100\,.
    \label{eq:eps_corr}
\end{equation}
Values for $m_1$ are displayed in Fig.~\ref{fig:int_corr_m1}. The IS0 and IS2 channels display similar effects, with a similar ordering of the four interactions and deviations between RPA and IMSRG results typically smaller than 5\% (except for lighter systems). This trend is strongly modified for the IV1 response, for which an up to 40\% increase is observed associated with many-body correlations. This effect is significantly smaller (15\%) for the \magicint{} compared to, \eg{}, the \nnlosat{} interaction (about 40 \%). Large deviations between RPA and IMSRG results in the IV1 sector signal the critical role of many-body correlations in a consistent description of photoabsortpion cross sections, which is detailed in the following sections.

In addition, we show the ratio $m_1/m_0$ in Fig.~\ref{fig:int_E}. The comparison of the results at the RPA and IMSRG level shows that many-body correlations tend to shift the energy centroid towards higher energy in the IS0 and IS2 channels, respectively by 2.2 and 2.7\,MeV. This shift is significantly enhanced in the IV1 channel, with an average of 9.8\,MeV. More interestingly the variation of the centroid energy for different interactions is strongly reduced in the IV1 and IS2 channels going from the RPA to the IMSRG, with a drop from 14 to 1.7\% and from 12 to 3.9\% in the IV1 and IS2 channels, respectively. 
The same trend, however, is not observed in the IS0 channel. The monopole response is more strongly related to compression properties of the underlying interaction, such that variations in the incompressibility of the different interactions (see Ref.~\cite{Alp25a}) impact the monopole response of finite nuclei. 

\begin{figure*}[t!]
    \centering
    \includegraphics[width=\textwidth]{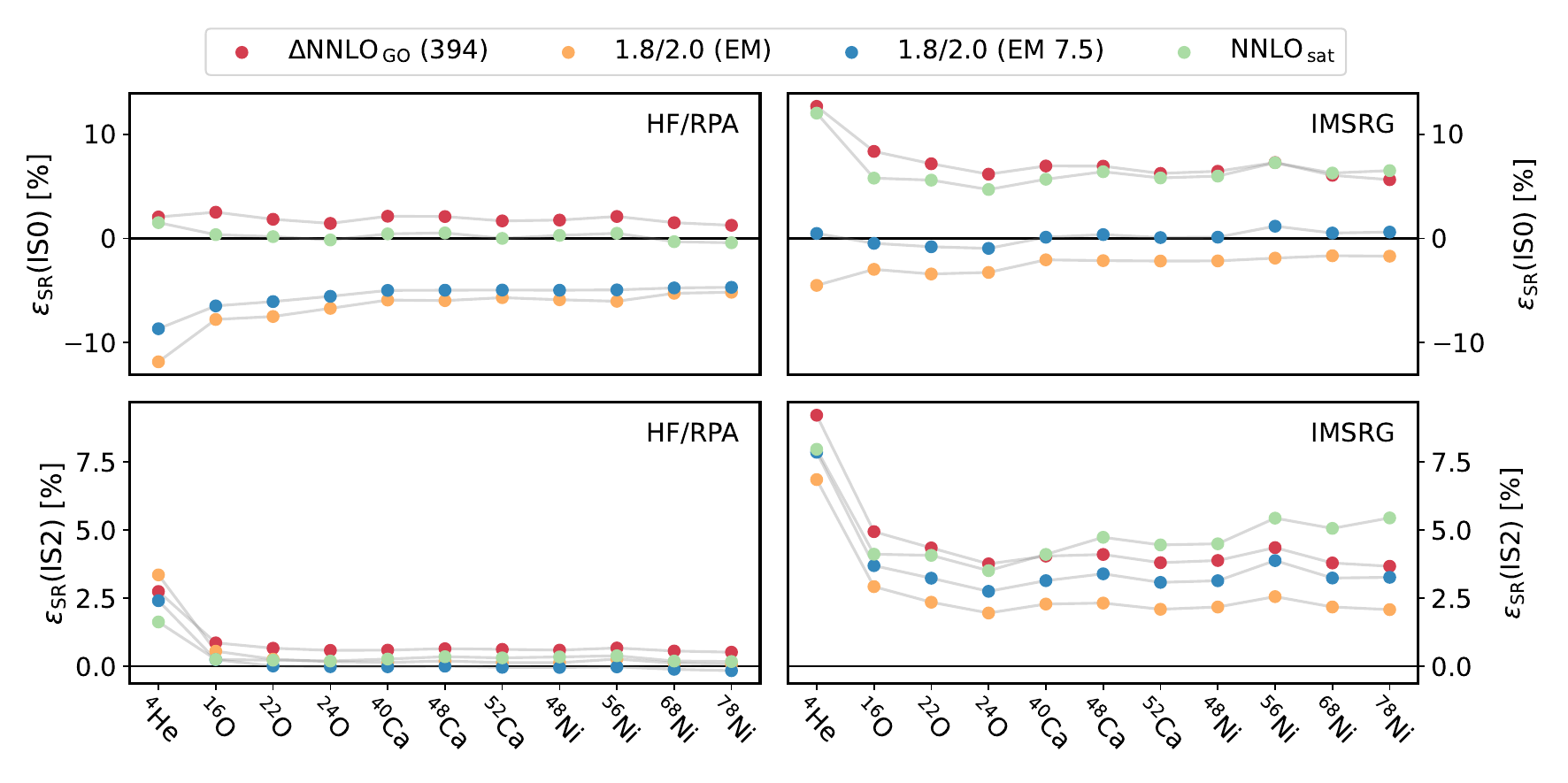}
    \caption{Relative difference between the EWSR and $m_1$, as defined by Eq.~\eqref{eq:eps_sr} at the HF/RPA (left panels) and IMSRG(2) (right panels) level, in the IS0 (top), IV1 (middle), and IS2 (bottom) channels. Results are given for four different interactions (\deltago{}~\cite{Jiang20a}, \magicint{}~\cite{Hebeler10a}, \emarthuis{}~\cite{Arthuis24a}, and \nnlosat{}~\cite{Ekstrom15a}) for $\Xm{e}=12$ and $\hbar\omega=$16\,MeV.}
    \label{fig:int_sr}
\end{figure*}

\subsection{Comparison to sum rules}
\label{sec:res_sr}

The difference between the EWSR and $m_1$ due to nonlocal interactions and the inclusion of two-body currents through the commutator expression in $m_1$ was discussed in Sec.~\ref{sec:sum_rules}. In this section the results obtained using the ESWRs defined in Eqs.~\eqref{eq:srs} and~\eqref{eq:trk} are compared to the values of $m_1$ obtained from the expectation value of $M_1(Q_\lambda)$, see Eq.~\eqref{eq:double_comm}. The EWSRs are evaluated, for the IS0 and IS2 operators, using the HF and IMSRG(2) expectation values of the intrinsic radius (Eqs.~\eqref{eq:srs}), while the IV1 sum rule is simply a number, see Eq.~\eqref{eq:trk}, and as such it is independent on the underlying nuclear state (which is why we do not explore the IV1 case in the following).

In order to quantitatively discuss the differences between the two approaches, we introduce the relative difference
\begin{equation}
\label{eq:eps_sr}
    \varepsilon_{\text{SR}}(Q_\lambda) [\%]\equiv\frac{m_1(Q_\lambda)-\text{EWSR}(Q_\lambda)}{\text{EWSR}(Q_\lambda)}\times 100 \,.
\end{equation}
Results are displayed in Fig.~\ref{fig:int_sr} for the IS0 and IS2 operators, at the HF/RPA (left panels) and IMSRG(2) (right panels) level. For the ISO response, we find a larger different between the EWSR and $m_1$ for the SRG-evolved interactions \magicint{} and \emarthuis{}. This is likely due to the increased nonlocality at lower SRG resolution scales. With correlations included at the IMSRG level, the relative difference between the EWSR and $m_1$ is larger for the \deltago{} and \nnlosat{} interactions, which are not as nonlocal as SRG-evolved interactions. We attribute this to the contributions from two-body currents included through the commutator expression for $m_1$.

In the bottom panels of Fig.~\ref{fig:int_sr} the same analysis shows that the difference is close to zero at the HF/RPA level for all the interactions considered, except for $^4$He where the different trend may be due to  an incomplete factorization of the cm motion.
At the IMSRG(2) level, we find a similar difference between the EWSR and $m_1$ for the IS2 for all interactions, at a similar level as for the IS0 channel. This is again due to the implicit inclusion of two-body currents in $m_1$. In general, this discussion shows that the $m_1$ response is more accurate than the EWSR expression.

\subsection{Enhancement factor of the TRK sum rule}

\begin{figure}[t!]
    \centering
    \includegraphics[width=1\columnwidth]{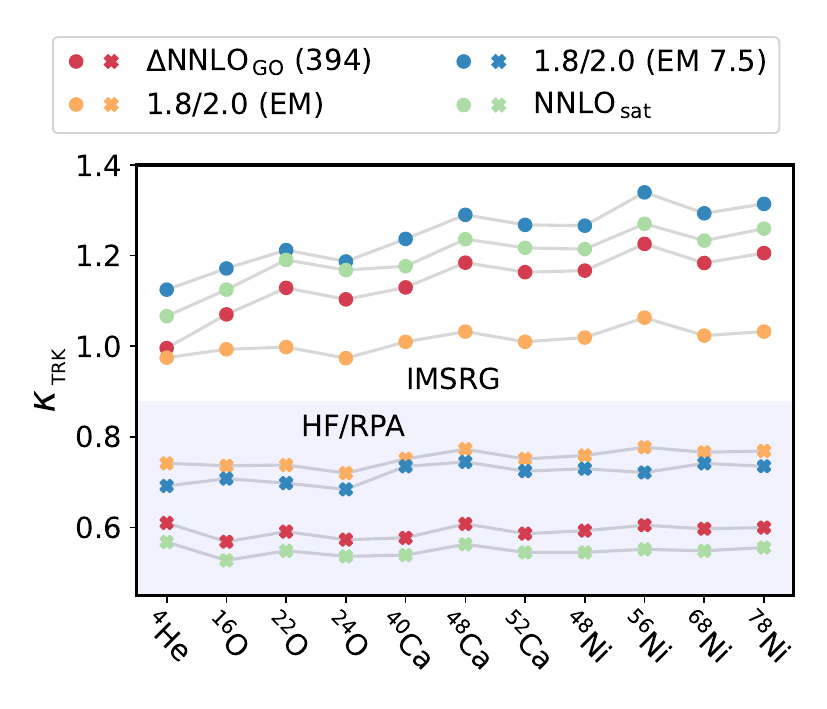}
    \caption{Thomas-Reike-Kuhn enhancement factor $\kappa_\text{TRK}$ at the HF/RPA and IMSRG(2) level. Results are given for four different interactions (\deltago{}~\cite{Jiang20a}, \magicint{}~\cite{Hebeler10a}, \emarthuis{}~\cite{Arthuis24a}, and \nnlosat{}~\cite{Ekstrom15a}) for $\Xm{e}=12$ and $\hbar\omega=$16\,MeV.}
    \label{fig:int_trk}
\end{figure}

Next, we study deviations from the TRK sum rule~\eqref{eq:trk} for the IV1 response. Our results at the HF/RPA and IMSRG(2) level are displayed in Fig.~\ref{fig:int_trk} in terms of the so-called TRK enhancement factor
\begin{equation}
    \kappa_\text{TRK}\equiv\frac{m_1(\text{IV}1)}{\text{EWSR}_{\text{int}}(E1)}-1\,,
\end{equation}
which is a common quantity used in nuclear physics to quantify deviations from the atomic analogue. Earlier works based on techniques similar to the ones employed here can be found in Refs.~\cite{Gari78a,Traini86a}. The magnitude of $\kappa_\text{TRK}$ is indicative of the presence of exchange terms in nuclear interactions. A very similar value for $\kappa_\text{TRK}$ is observed for all nuclei studied at the uncorrelated HF level (lower half of the plot) independently of the interaction used. The \deltago{} and \nnlosat{} interactions give values of $\kappa_\text{TRK}$ around 0.6, increasing to 0.8 for the SRG-evolved interactions \magicint{} and  \emarthuis{}. We note that the results found here at the HF/RPA level with \nnlosat{} are consistent with the ones observed in RPA calculations employing the same Hamiltonian~\cite{Wu18a}. It is interesting to see that, already at the uncorrelated HF level, chiral interactions give values of $\kappa_\text{TRK}$ systematically larger compared to EDF calculations, a point that will be discussed in Sec.~\ref{sec:exp}. This reflects the presence of exchange terms within chiral interactions. After the IMSRG evolution a further enhancement of $\kappa_\text{TRK}$ is observed due to many-body correlations and the inclusion of two-body currents in $m_1$, with a moderate tendency to increase for the heavier nuclei studied.

\begin{figure*}[t!]
    \centering
    \includegraphics[width=\textwidth]{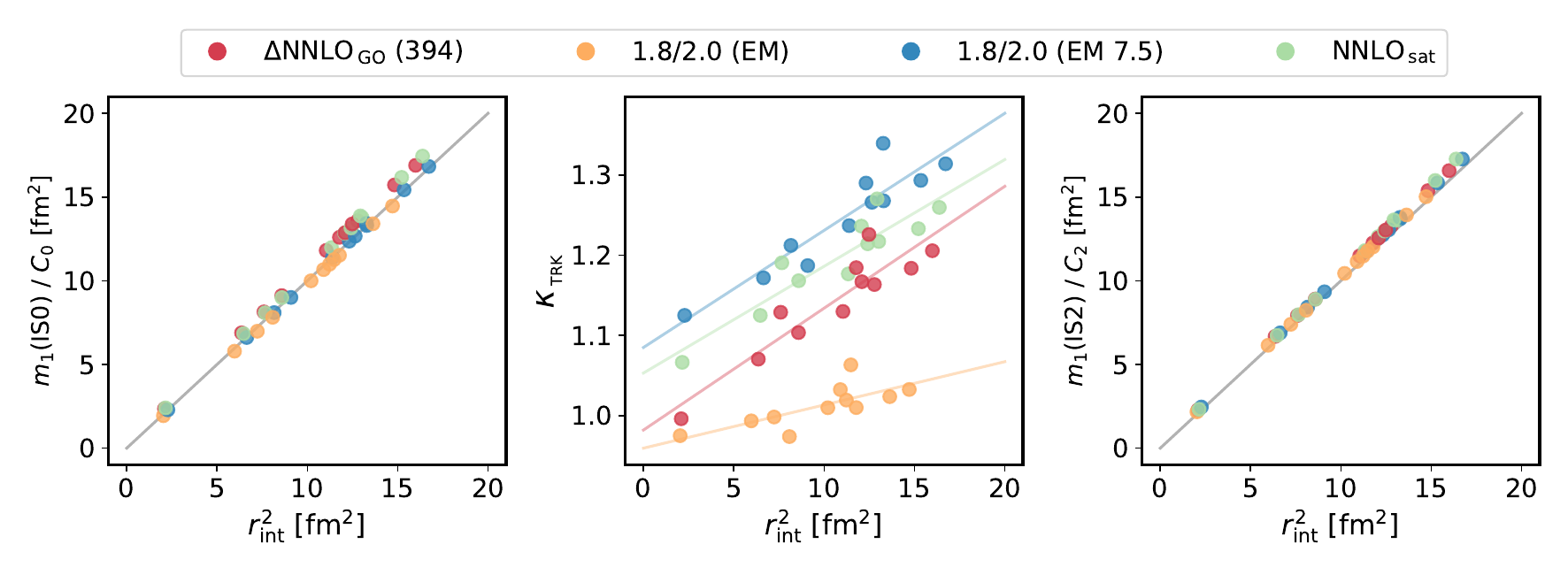}
    \caption{Correlation of $m_1$ with the intrinsic mean square radius for the IS0 ($m_1/C_0$, left panel), IV1 ($\kappa_\text{TRK}$, middle panel), and IS2 ($m_1/C_2$, right panel) operators at the IMSRG(2) level. Results are given for four different interactions (\deltago{}~\cite{Jiang20a}, \magicint{}~\cite{Hebeler10a}, \emarthuis{}~\cite{Arthuis24a}, and \nnlosat{}~\cite{Ekstrom15a}) for $\Xm{e}=12$ and $\hbar\omega=$16\,MeV.}
    \label{fig:correlation_plot}
\end{figure*}

In order to probe the correlations with the intrinsic mean square radius, the IMSRG(2) values for $m_1$ are plotted in Fig.~\ref{fig:correlation_plot} versus $\braket{r^2_\text{int}}$. The results for the IS0 and IS2 operators are rescaled following Eqs.~\eqref{eq:srs} by the factors
\begin{subequations}
    \begin{align}
        C_0&\equiv\frac{2\hbar^2}{m}\,,\\
        C_2&\equiv\frac{25}{4\pi}\frac{\hbar^2}{m} \,.
    \end{align}    
\end{subequations}
Although deviations were observed in Fig.~\ref{fig:int_sr}, there is a clear linear correlation, independent of the nucleus or the underlying interaction. For the IV1 response, this is studied in the middle panel of Fig.~\ref{fig:int_sr}, where the TRK enhancement factor $\kappa_\text{TRK}$ is plotted versus $\braket{r^2_\text{int}}$. Linear fits of the form
\begin{equation}
    \kappa_\text{TRK} = a\braket{r^2_\text{int}}+b\,,
    \label{eq:lin_fit}
\end{equation}
are also plotted for the different interactions, with the best values for $a$ and $b$ reported in Tab.~\ref{tab:fit_coeff}. We only find a weak correlation in this case for each interaction, with similar slopes for the \emarthuis{}, \nnlosat{}, and \deltago{} interactions.

\begin{table}[t!]
    \renewcommand{\arraystretch}{1.2}
    \begin{tabular*}{\linewidth}{@{\extracolsep{\fill}}l cc}  
    \hline\hline
    & {$a$ [fm$^{-2}$]} & {$b$} \\
    \hline
    \deltago{} & 0.015(2) & 0.98(2) \\
    \magicint{} & 0.005(2) & 0.96(2) \\
    \emarthuis{} & 0.015(2) & 1.08(2) \\
    \nnlosat{} & 0.013(2) & 1.05(2) \\
    \hline\hline
    \end{tabular*}
\caption{Fitting coefficients for the linear extrapolation in the middle panel of Fig.~\ref{fig:correlation_plot}, as given by Eq.~\eqref{eq:lin_fit}.}
\label{tab:fit_coeff}
\end{table}

\begin{figure*}
    \centering
    \includegraphics[width=0.49\linewidth]{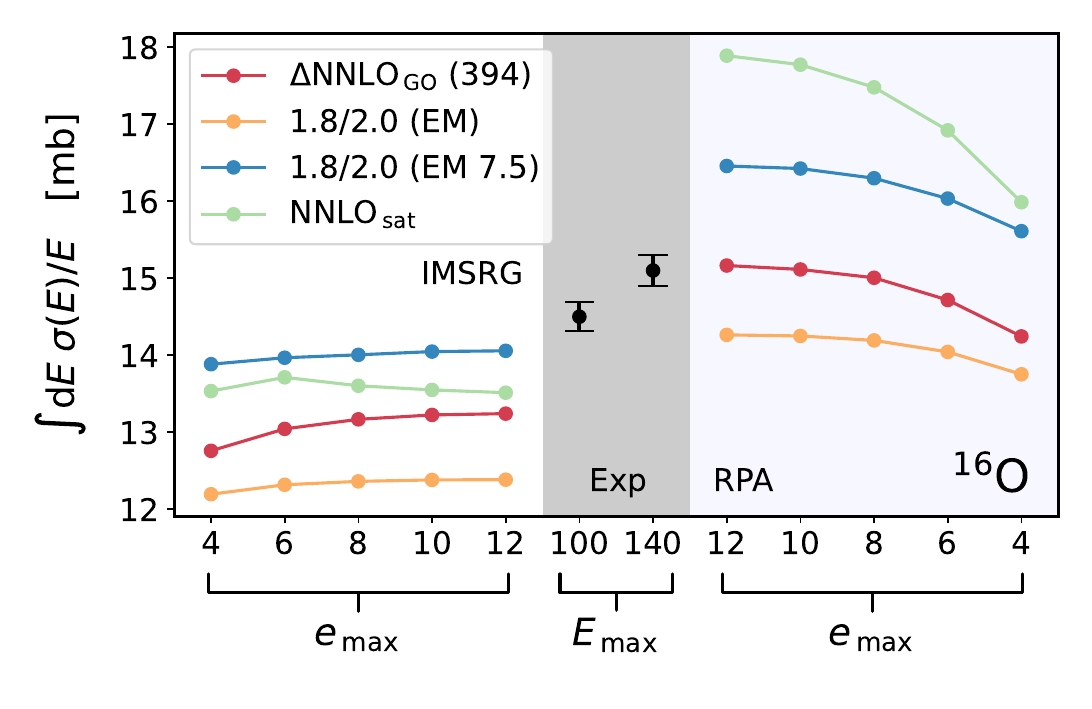}
    \includegraphics[width=0.49\linewidth]{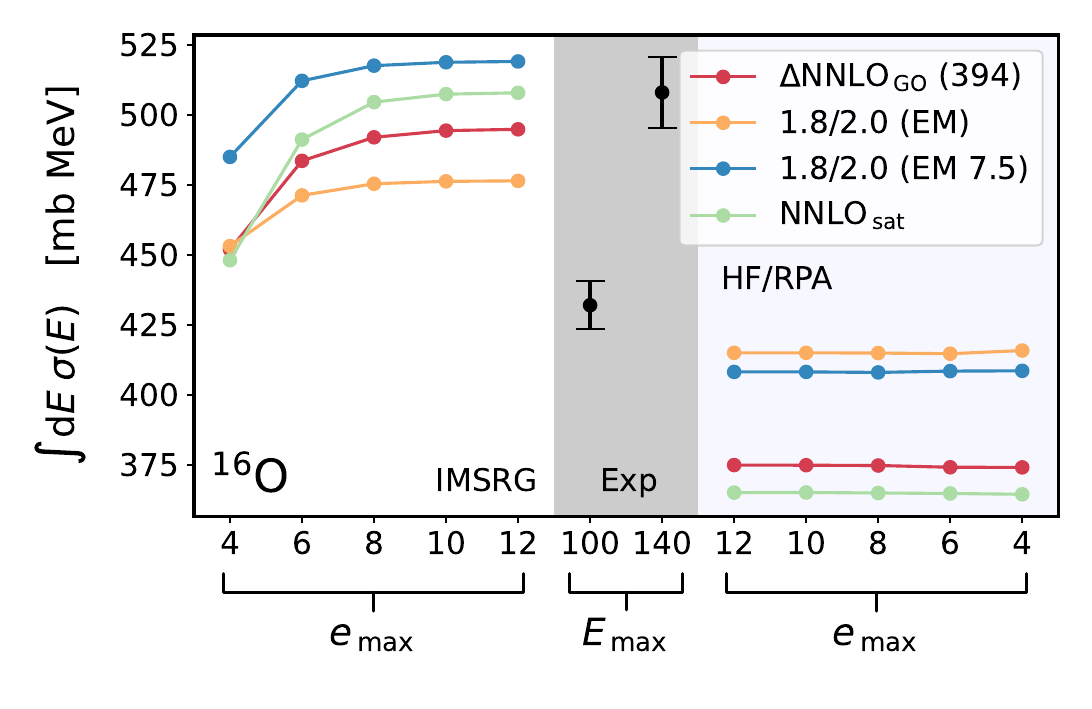}
    \caption{Integrated photoabsorption cross sections for $^{16}$O. The left panel is weighed by $1/E$. Experimental values for different maximum energies are taken from Ref.~\cite{Ahrens75a}. Our results at the IMSRG(2) (left) and RPA level (right in each panel) are shown for the four different interactions (\deltago{}~\cite{Jiang20a}, \magicint{}~\cite{Hebeler10a}, \emarthuis{}~\cite{Arthuis24a}, and \nnlosat{}~\cite{Ekstrom15a}) with increasing $\Xm{e}$ up to $12$ and for $\hbar\omega=$16\,MeV.}
    \label{fig:O16_IV1}
\end{figure*}

\begin{figure*}
    \centering
    \includegraphics[width=0.49\linewidth]{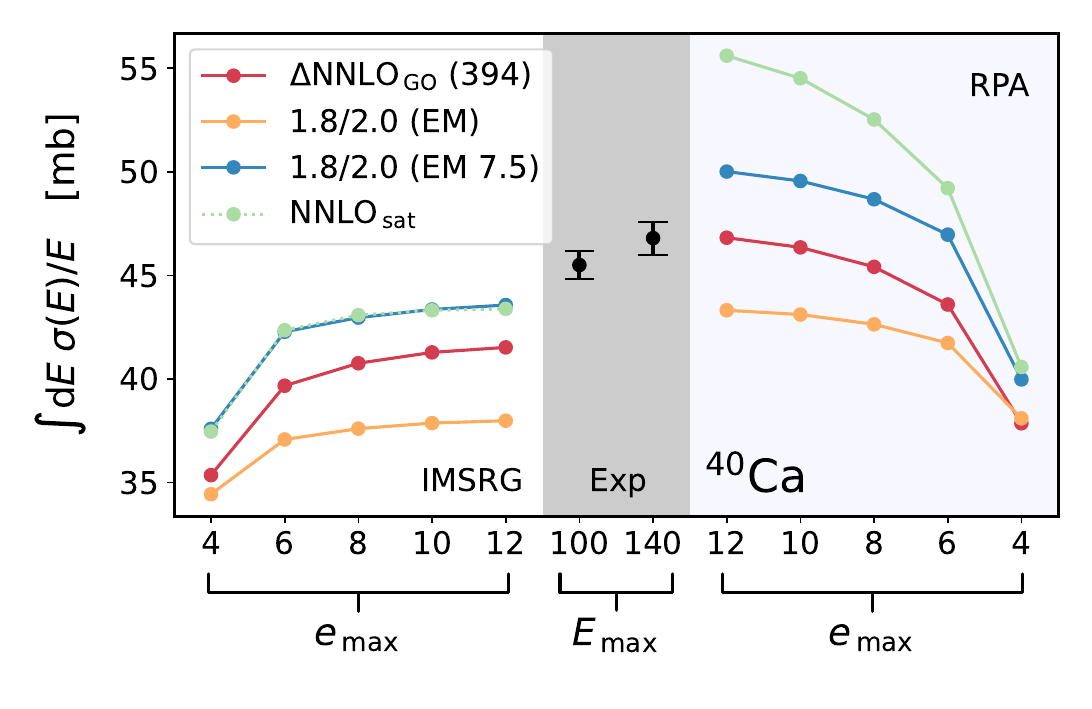}
    \includegraphics[width=0.49\linewidth]{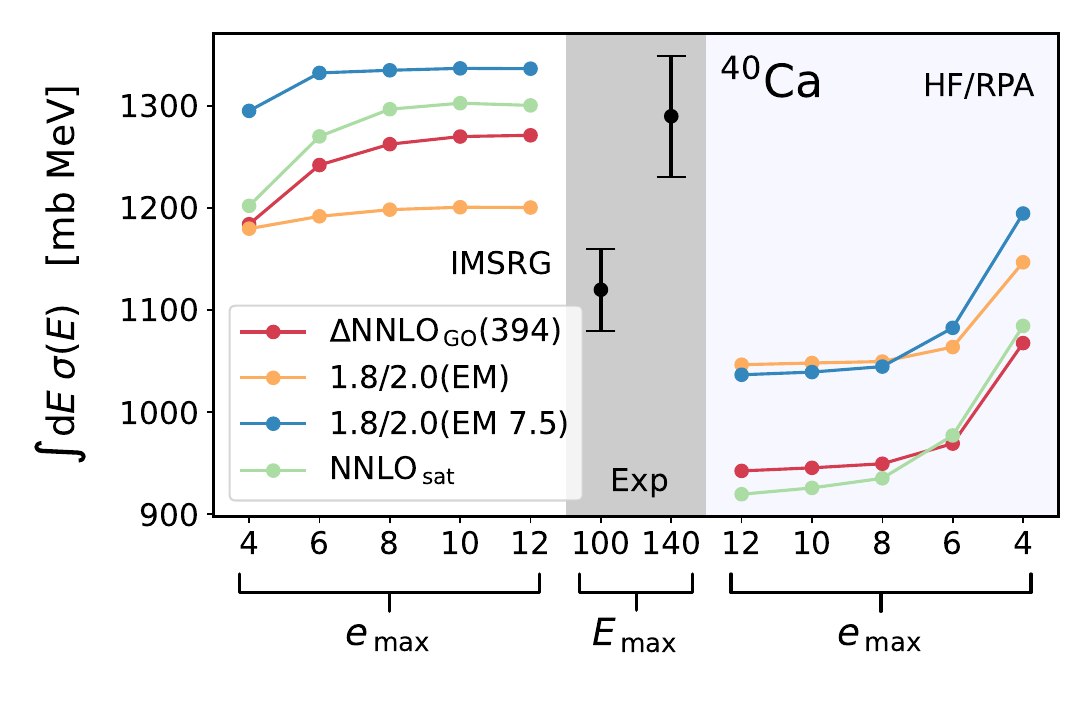}
    \caption{Same as Fig.~\ref{fig:O16_IV1} but for the integrated photoabsorption cross section for $^{40}$Ca. The experimental values for different maximum energies are also taken from Ref.~\cite{Ahrens75a}.}
    \label{fig:Ca40_IV1}
\end{figure*}

\subsection{Comparison to experiment: dipole response}
\label{sec:exp}

Experiments have extensively explored the multipole response of nuclei, see the reviews
addressing the IS0~\cite{Garg18a}, the IV1~\cite{Bracco19a}, and the IS2 response~\cite{Bertrand81a}. However, the experimental efforts have mostly focused on the giant resonance region. Thus, data have been collected up to an excitation energy rarely larger than 40\,MeV, while a consistent comparison to the theoretical predictions of this work requires integration up to high excitation energies. To the best of our knowledge, such an effort has only been made for the IV1 response in selected nuclei. As the dipole channel represents the dominant contribution to the total electric response, the total photoabsorption cross section can be directly used. For the set of nuclei studied in the present work, only $^{16}$O and $^{40}$Ca have been studied experimentally~\cite{Ahrens75a} up to the pion-production threshold. Results from Ref.~\cite{Ahrens75a} came as a surprise at the time, as the extracted value of the TRK enhancement factor was underpredicted by theoretical models~\cite{Traini86a}.

In this section, we compare the integrated cross sections\footnote{The prefactor in Eqs.~\eqref{eq:cross_section} differs by the usual one, see, \eg, Refs.~\cite{Orlandini91a,Ring1980_ManyBodyBook}, because all components of the dipole operator are used here, together with the Wigner-Eckart theorem, instead of the $z$-component only.}
\begin{subequations}
\label{eq:cross_section}
    \begin{align}
        \int dE\,\sigma(E)&=\frac{16\pi^3}{9}\alpha\,m_1(\text{IV}1)\,,\\
        \int dE\,\frac{\sigma(E)}{E}&=\frac{16\pi^3}{9}\alpha\,m_0(\text{IV}1)\,,\label{eq:brem}
    \end{align}
\end{subequations}
where $\alpha$ is the fine-structure constant, to the experimental data from Ref.~\cite{Ahrens75a} for $^{16}$O and $^{40}$Ca. Equation~\eqref{eq:brem} is also referred to as the bremsstrahlung sum rule in the literature~\cite{Orlandini91a,Gazit06a}. Experimental results are shown for integration up to two maximum energies $E_\text{max}$ of 100 and 140\,MeV.

Our HF/RPA and IMSRG(2) results for $^{16}$O are compared to the the experimental values in Fig.~\ref{fig:O16_IV1}. In the left panel, a consistent reduction in the spread between different interactions is observed in going from the HF/RPA to the IMSRG(2) description. The IMSRG results underpredict the experimental value by 6 to 20\%, depending on the considered interaction. The agreement between the theoretical predictions and the experimental values is enhanced when discussing the integrated cross section (right panel of Fig.~\ref{fig:O16_IV1}). In this case the the IMSRG(2) results are in excellent agreement with the measured value for the larger maximum energy, except for the \magicint{} interaction, while HF/RPA underpredicts experiment by roughly 20\%. Completely analogous observations hold for $^{40}$Ca, for which results are displayed in Fig.~\ref{fig:Ca40_IV1}. The better agreement of the integrated cross section is a further confirmation of the higher quality of the $m_1$ predictions, with respect to $m_0$, as discussed in Sec.~\ref{sec:ms_conv}. 

\begin{figure}[t!]
    \centering
    \includegraphics[width=\columnwidth]{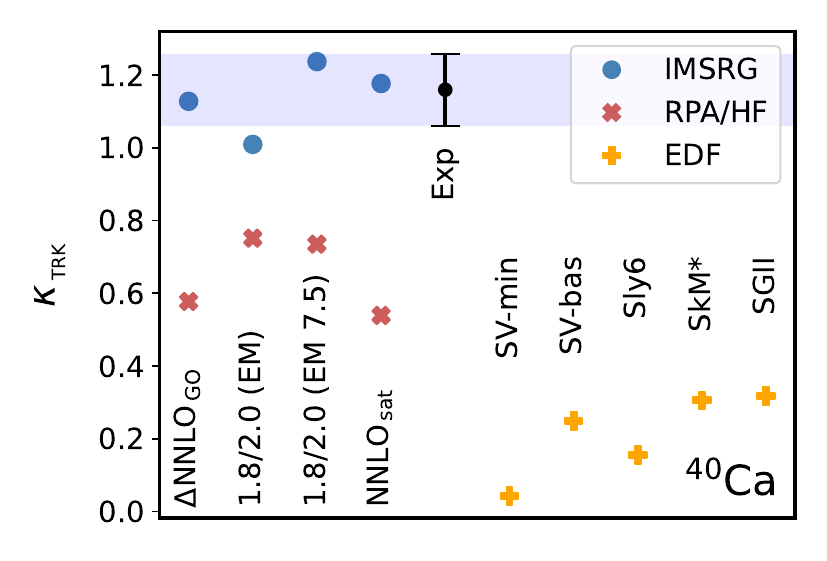}
    \caption{Values of the TRK enhancement factor $\kappa_\text{TRK}$ in $^{40}$Ca from our HF/RPA and IMSRG(2) calculations, compared to the experimental value from Ref.~\cite{Ahrens75a} (obtained from the maximum energy available corresponding to the pion-production threshold) and to Skyrme-based EDF calculations~\cite{Reinhard25a}.}
    \label{fig:Ca40_trk}
\end{figure}

The integrated photoabsorption cross section allows to extract the TRK enhancement factor $\kappa_\text{TRK}$. This is shown in Fig.~\ref{fig:Ca40_trk} for our HF/RPA and IMSRG(2) calculations compared to the extracted experimental value~\cite{Ahrens75a} and to Skyrme-based RPA calculations~\cite{Reinhard25a}. 
For the different Skyrme parametrizations, EDF calculations give a too low enhancement factor between about $0.1-0.35$, significantly underestimating the experimental value of $\kappa_\text{TRK} \approx 1.15$ for \elem{Ca}{40}.
Already at the HF/RPA level, the different interactions considered yield larger values of $0.5-0.8$. Including many-body correlations at the IMSRG(2) level (and two-body currents through the commutator expression) provides a very good description of $\kappa_\text{TRK}$ for all interactions.
The difference between mean-field and IMSRG(2) calculations may be explained by the presence of correlated neutron–proton pairs (quasi-deuterons). These correlated pairs are responsible for most of the photoabsorption cross section for incident photons in the energy range $40–140\,$MeV~\cite{Levinger51a,Chadwick91a,Goriely20a}, and are taken into account at the IMSRG(2) but not at the HF level or in mean-field calculations. Further studies, however, are in order for a more quantitative statement in this direction.

\subsection{Valence-space extension}

So far many-body calculations employed the single-reference formulation of the IMSRG that is applicable to closed-shell nuclei.
A widely used extension of the IMSRG is given by its valence-space (VS) formulation, where a VS Hamiltonian on top of a core is decoupled from excitations outside the VS~\cite{Tsukiyama12a,Bogner14a,Stroberg16a,Hergert16a,Stroberg19a,Miyagi20a}. The VS Hamiltonian is subsequently diagonalized to access the ground state and low-lying spectroscopy.
In this work, we probe the effect of the VS description by comparing results from the single-reference IMSRG to the VS-IMSRG using a \elem{Si}{28} core. 
Results for the ground-state energy and mean square matter radius are given in Tab.~\ref{tab:VS_diff} for the four different interactions.
In all cases, the VS-IMSRG calculations give lower ground-state energies by about $7-10 \, \MeV$, \ie{}, a difference of $2-3\%$ to the total ground-state energy.
On the other hand, the mean square matter radii are smaller in the VS-IMSRG by $4-6\, \fm^2$, which amounts to a decrease by $1-1.5\%$.
The too small radius of the \magicint{} interaction is well known and has been recently addressed through the design of the \emarthuis{} interaction.
The \nnlosat{} interaction gives a slightly lower ground-state energy, which we attribute to the missing triples corrections for this somewhat harder interaction.

\begin{table}[t!]
    \renewcommand{\arraystretch}{1.2}
    \begin{tabular*}{\linewidth}{@{\extracolsep{\fill}}l cccc}  
    \hline\hline
    & \multicolumn{2}{c}{$E$ [MeV]}   & \multicolumn{2}{c}{$A \braket{r^2_\text{int}}$ [fm$^2$]} \\
    & single-ref. & $^{28}$Si & single-ref. & $^{28}$Si \\ \hline
    \deltago{} & $-339.66$ & $-349.69$ & 447.22 & 441.25 \\
    \magicint{} & $-344.54$ & $-351.10$ & 413.52 & 409.19 \\
    \emarthuis{} & $-341.93$ & $-351.29$ & 461.10 & 457.04 \\
    \nnlosat{} & $-326.71$ & $-335.03$ & 459.41 & 453.97 \\
    \hline\hline
    \end{tabular*}
\caption{Ground-state energies and mean square matter radii of \elem{Ca}{40} for the interactions considered. Observables are either evaluated using the single-reference IMSRG(2) or the valence-space formulation with a \elem{Si}{28} core.}
\label{tab:VS_diff}
\end{table}

\begin{figure}[t!]
    \centering
    \includegraphics[width=\columnwidth]{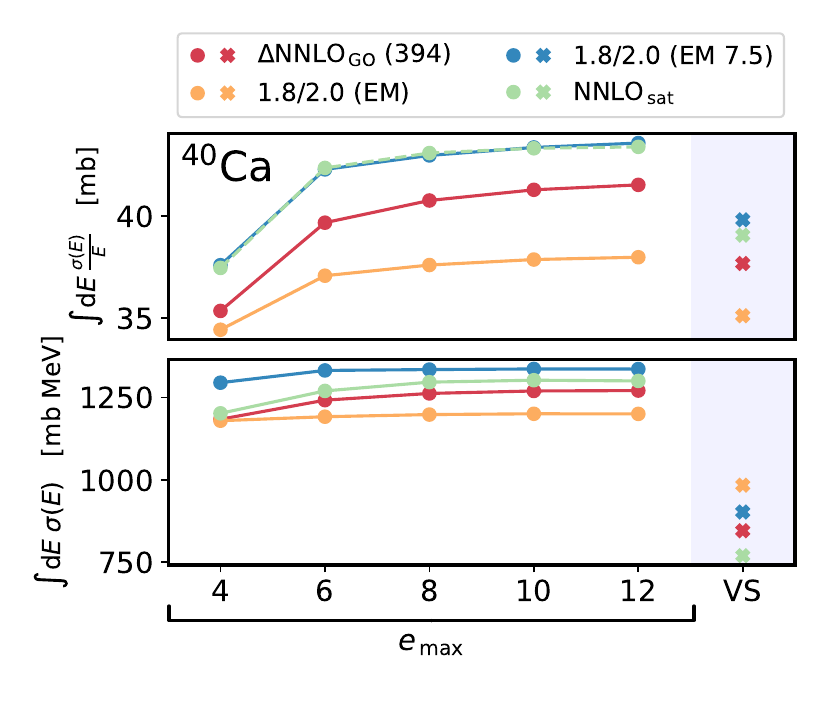}
    \caption{Integrated cross sections from single-reference IMSRG calculations at different basis sizes \Xm{e} compared to valence-space (VS) IMSRG calculations based on a $^{28}$Si core ($\Xm{e}=12$). Results are shown for the four different interactions for $\hbar\omega=$16\,MeV.}
    \label{fig:Ca40_VS}
\end{figure}

The differences between the VS and single-reference descriptions of the nuclear response are shown in Fig.~\ref{fig:Ca40_VS}. For both integrated cross sections, we find significantly smaller values with the VS-IMSRG. This could be due to the two-step decoupling process employed in the VS-IMSRG, which is an additional source of uncertainty. This needs to be studied more in the future using also other multi-reference extensions of the IMSRG.

\section{Conclusions}
\label{sec:concl}

In this work, we developed a framework that enables \ai{} calculations of integrated properties of the nuclear response. Using the IMSRG, we studied the multipole response of nuclei based on chiral NN and 3N interactions, with a particular focus on the impact of ground-state correlations. We investigated the isoscalar mono- and quadrupole as well as the isovector dipole response of closed-shell nuclei from $^4$He to $^{78}$Ni. Our results show that the inclusion of many-body correlations yields significant contributions compared to the HF/RPA level and yields much better agreement with available experimental data in \elem{O}{16} and \elem{Ca}{40}. In particular, our calculations provide a very good description of the TRK enhancement factor in \elem{O}{16} and \elem{Ca}{40}, which is systematically underpredicted in mean-field calculations. We also discussed how the moment method takes into account many-body currents through the commutator expression, which is thus superior to simple evaluations of the EWSRs implicitly assuming the interaction to be local.

Our results demonstrate that the moment method can be used as a benchmark for other \ai{} approaches that describe nuclear response functions through the explicit treatment of excited states. In addition, we performed first valence-space IMSRG studies to extend this work top open-shell nuclei. Our results for \elem{Ca}{40} showed that this needs further studies and also benchmarking with other multi-reference methods for open-shell nuclei. The advantage of the moment method is that this is straightforward to use in other \ai{} approaches, as it is only based on the correlated ground state and does not require the inclusion of excited states.

\section*{Data availability}
The data that support the findings of this article are openly available~\cite{zenodo}.

\section*{Acknowledgements}

We thank P.-G. Reinhard for sharing EDF results for the TRK enhancement factor as well as for insightful discussions. Discussions with G. Col\`o, M. Drissi, T. Duguet, X. Roca-Maza, and V. Som\`a are also gratefully acknowledged. The work of A.P. and A.S. is supported by the Deutsche Forschungsgemeinschaft (DFG, German Research Foundation) -- Project-ID 279384907 -- SFB 1245. A.T. is supported by the European Research Council (ERC) under the European Union's Horizon Europe research and innovation programme (Grant Agreement No.~101162059). The authors gratefully acknowledge the Gauss Centre for Supercomputing e.V. (www.gauss-centre.eu) for funding this project by providing computing time on the GCS Supercomputer JUWELS at J\"ulich Supercomputing Centre (JSC).

\appendix

\section{Intrinsic sum rules}
\label{sec:app_intrinsic}

The EWSRs for isoscalar local operators are textbook material. To the best of our knowledge their derivation is mostly limited to the laboratory frame. The derivation for the intrinsic monopole operator can be found in Ref.~\cite{Porro24c}. In this appendix, a derivation of the EWSR for isoscalar operators in the intrinsic frame is generalized to all multipole operators. The excitations under consideration are intrinsic ones, because the nuclear interaction excites the internal degrees of freedom of the system. Thus, excitation operators should be expressed in the intrinsic frame to avoid spurious contributions from excitations of the center of mass. Upon introduction of the cm position and momentum vectors
\begin{subequations}
    \begin{align}
        \vec{R}&=\frac{1}{A}\sum_{i=1}^A\vec{r}_i\,,\\
        \vec{P}&=\sum_{i=1}^A\vec{p}_i\,,
    \end{align}
\end{subequations}
the particle coordinates in the intrinsic frame are 
\begin{subequations}
    \begin{align}
        \vec{\xi}_i&=\vec{r}_i-\vec{R}\,,\\
        \vec{\pi}_i&=\vec{p}_i-\frac{\vec{P}}{A}\,.
    \end{align}
\end{subequations}
These newly introduced variables satisfy slightly modified canonical commutation relations
\begin{equation}
    [\vec{\pi}_k,\vec{\xi}_l]=[\vec{p}_k,\vec{r}_l]-\frac{3i\hbar}{A}\,,
\end{equation}
where the second term nicely shows how the change of coordinates acts like a projection operator that removes the cm component. It is immediate to check that the following equalities hold 
\begin{equation}
    [\vec{\pi}_k,\vec{\xi}_l] = [\vec{p}_k,\vec{\xi}_l] = [\vec{\pi}_k,\vec{r}_l]\,,
\end{equation}
meaning that the projection of one coordinate automatically takes care of removing the cm component from the conjugate variable as well. Therefore, when evaluating the intrinsic EWSR it is not necessary to express both the Hamiltonian and the excitation operator in the intrinsic frame. Expressing either one in the intrinsic frame is sufficient to eliminate the cm contribution from the other.

In this work, the Hamiltonian is formulated in the intrinsic frame, while the multipole operators are retained in the laboratory frame. The kinetic energy reads
\begin{equation}
    T_\text{int}=\frac{1}{2m}\sum_{i=1}^Ap_i^2-\frac{1}{2mA}\sum_{i,j=1}^A\vec{p}_i\cdot\vec{p}_j\,,
\end{equation}
such that one has to evaluate the quantity
\begin{equation}
\label{eq:dim_1}
    \text{EWSR}_{\text{int}}(Q_{\lambda})\equiv\frac{1}{2}\!\sum_{\mu=-\lambda}^{\lambda}\!\braket{\Psi_0|[Q_{\lambda\mu}^\dagger,[T_\text{int},Q_{\lambda\mu}]]|\Psi_0} \,.
\end{equation}
We start from the general expression
\begin{align}
\label{eq:gen_sr_int}
    \frac{1}{2}[F,[T_\text{int},G]]\,,
\end{align}
with $F$ and $G$ isoscalar one-body operators of the form
\begin{equation}
    F=\sum_{i=1}^Af(\vec{r}_i)\,.
\end{equation}
Equation~\eqref{eq:gen_sr_int} is expanded using
\begin{equation}
    [\vec{p}_i,f(\vec{r}_k)]=-i\hbar\vec{\nabla}_if(\vec{r}_i)\delta_{ik}\,,
\end{equation}
and the general contributions have the structure
\begin{align}
    \frac{1}{2}[F,[\vec{p}_i\cdot\vec{p}_j,G]]&=\frac{1}{2}\sum_{kl=1}^A[f(\vec{r}_k),[\vec{p}_i\cdot\vec{p}_j,g(\vec{r}_l)]]\nonumber\\
    &=-\sum_{kl=1}^A[\vec{p}_i,f(\vec{r}_k)]\cdot[\vec{p}_j,g(\vec{r}_l)]\nonumber\\
    &=\hbar^2\vec{\nabla}_if(\vec{r}_i)\cdot\vec{\nabla}_jg(\vec{r}_j)\,.
\end{align}
In this work the functions are of the form $f(\vec{r})=r^\lambda Y_{\lambda\mu}(\hat{r})$, such that the action of the gradient gives, for each piece~\cite{Varshalovich88a},
\begin{equation}
    \vec{\nabla}[r^\lambda Y_{\lambda\mu}(\hat{r})]=\sqrt{\lambda(2\lambda+1)}r^{\lambda-1}\vec{Y}_{\lambda\mu}^{\lambda-1}(\hat{r})\,,
\end{equation}
where $\vec{Y}_{\lambda\mu}^{L}(\hat{r})$ are vector spherical harmonics. Summing up one can write, for the two contributions of the intrinsic kinetic energy,
\begin{widetext}
\begin{subequations}
    \begin{align}
        \frac{1}{2}\sum_{i=1}^A\sum_{\mu=-\lambda}^{\lambda}[Q_{\lambda\mu}^\dagger,[\frac{p_i^2}{2m},Q_{\lambda\mu}]]&=\frac{\hbar^2\lambda(2\lambda+1)}{2m}\sum_{\mu=-\lambda}^\lambda\sum_{i=1}^A r_i^{2\lambda-2}\vec{Y}_{\lambda\mu}^{\lambda-1\,*}(\hat{r}_i)\cdot\vec{Y}_{\lambda\mu}^{\lambda-1}(\hat{r}_i)\,,\label{eq:dim_part1}\\
        \frac{1}{2}\sum_{ij=1}^A\sum_{\mu=-\lambda}^{\lambda}[Q_{\lambda\mu}^\dagger,[\frac{\vec{p}_i\cdot\vec{p}_j}{2mA},Q_{\lambda\mu}]]&=\frac{\hbar^2\lambda(2\lambda+1)}{2mA}\sum_{\mu=-\lambda}^\lambda\sum_{ij=1}^A r_i^{\lambda-1}r_j^{\lambda-1}\vec{Y}_{\lambda\mu}^{\lambda-1\,*}(\hat{r}_i)\cdot\vec{Y}_{\lambda\mu}^{\lambda-1}(\hat{r}_j)\,.\label{eq:dim_part2}
    \end{align}
\end{subequations}
\end{widetext}
Eventually, the relations (see Eqs.~(81) and (102) from Sec.~7.3 of Ref.~\cite{Varshalovich88a})
\begin{subequations}
    \begin{align}
        \sum_{\mu=-\lambda}^\lambda\vec{Y}_{\lambda\mu}^{L_1\,*}(\hat{r})\cdot\vec{Y}_{\lambda\mu}^{L_2}(\hat{r})&=\delta_{L_1L_2}\frac{2\lambda+1}{4\pi}\,,\\
        \sum_{\mu=-\lambda}^\lambda\vec{Y}_{\lambda\mu}^{L_1\,*}(\hat{r}_1)\cdot\vec{Y}_{\lambda\mu}^{L_2}(\hat{r}_2)&=\delta_{L_1L_2}\frac{2\lambda+1}{4\pi}P_{L_1}(\cos\omega_{12})\,,
    \end{align}
\end{subequations}
are used to rewrite the sum rule contributions as
\begin{subequations}
\label{eq:sr_int}
\begin{align}
    \eqref{eq:dim_part1}&=\frac{\hbar^2}{2m}\frac{\lambda(2\lambda+1)^2}{4\pi}\sum_{i=1}^Ar_i^{2\lambda-2}\,,\label{eq:sr_lab}\\
    \eqref{eq:dim_part2}&=\frac{\hbar^2}{2mA}\frac{\lambda(2\lambda+1)^2}{4\pi}\sum_{ij=1}^Ar_i^{\lambda-1}r_j^{\lambda-1}P_{\lambda-1}(\cos\omega_{ij})\,.\label{eq:sr_corr}
\end{align}
\end{subequations}
In Eqs.~\eqref{eq:sr_int} $P_L(x)$ are Legendre polynomials of degree $L$ and 
\begin{equation}
    \cos\omega_{12}\!=\cos\vartheta_1\cos\vartheta_2\!+\sin\vartheta_1\!\sin\vartheta_2\cos(\varphi_1-\varphi_2)\,.
\end{equation}

Equation~\eqref{eq:sr_lab} is the EWSR in the laboratory frame, as it can be found, for instance, in Eq.~(8.159) of Ref.~\cite{Ring1980_ManyBodyBook}. The expressions are equal up to an additional factor $(2\lambda+1)$, which is due to the sum over the $\mu$ components of the spherical tensor in this work. Equation~\eqref{eq:sr_corr} is a correction which must be subtracted when working in the intrinsic frame. Replacing $\lambda=2$ into Eq.~\eqref{eq:sr_corr} one obtains the correction to the quadrupole EWSR
\begin{equation}
    \frac{\hbar^2}{mA}\frac{25}{4\pi}\sum_{ij=1}^Ar_ir_j\cos\omega_{ij}=\frac{\hbar^2}{mA}\frac{25}{4\pi}\sum_{ij=1}^A\vec{r}_i\cdot\vec{r}_j\,,
\end{equation}
which subtracted from the laboratory frame part [Eq.~\eqref{eq:sr_lab}]
\begin{equation}
\label{eq:r_int}
    \text{EWSR}_{\text{lab}}(Q_2)=\frac{\hbar^2}{m}\frac{25}{4\pi}\braket{\Psi_0|r^2|\Psi_0}
\end{equation}
gives
\begin{equation}
    \text{EWSR}_{\text{int}}(Q_2)=\frac{\hbar^2}{m}\frac{25}{4\pi}\braket{\Psi_0|r_{\text{int}}^2|\Psi_0} \,,
\end{equation}
where
\begin{subequations}
    \begin{align}
        r^2&\equiv\sum_{i=1}^Ar_i^2\,,\\
        r_{\text{int}}^2&\equiv\sum_{i=1}^Ar_i^2-\frac{1}{A}\sum_{ij=1}^A\vec{r}_i\cdot\vec{r}_j\,,
    \end{align}
\end{subequations}
so the two expressions are strictly equivalent upon replacement of the mean square radius in the laboratory frame with its intrinsic counterpart.

It is also interesting to show that this technique can be used to derive the TRK sum rule, which is usually obtained starting from the definition of the intrinsic dipole operator~\eqref{eq:dipole_int}. Here, instead, we can start from the electric dipole operator
\begin{equation}
    Q_{1\mu}^E=e\sum_{i=1}^Zr_iY_{1\mu}(\hat{r}_i)\,.
\end{equation}
Simply putting $\lambda=1$ into Eqs.~\eqref{eq:sr_int} and summing only over $Z$, the result
\begin{align}
    \text{EWSR}_{\text{int}}(Q_1)&=\frac{\hbar^2e^2}{2m}\frac{9}{4\pi}(Z-\frac{Z^2}{A})\nonumber\\
    &=\frac{\hbar^2e^2}{2m}\frac{9}{4\pi}\frac{NZ}{A}
\end{align}
is readily obtained, which is exactly the TRK sum rule fo the intrinsic dipole operator, see Eq.~\eqref{eq:dipole_int}.

\begin{figure*}
    \centering
    \includegraphics[width=\textwidth]{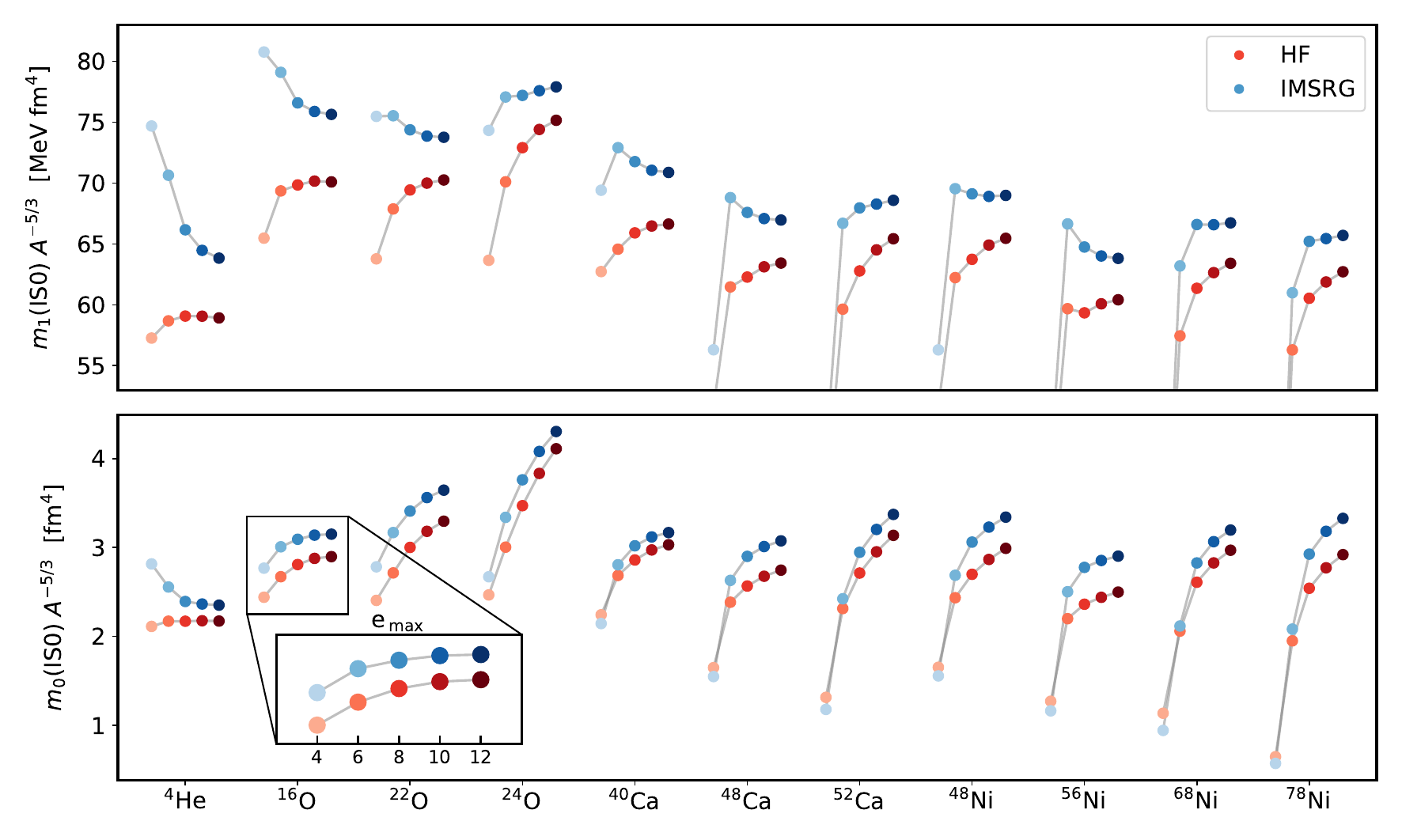}
    \caption{Same as Fig.~\ref{fig:L1_emax_conv} but for the isoscalar monopole (IS0) response.}
    \label{fig:L0_emax_conv}
\end{figure*}

\begin{figure*}
    \centering
    \includegraphics[width=\textwidth]{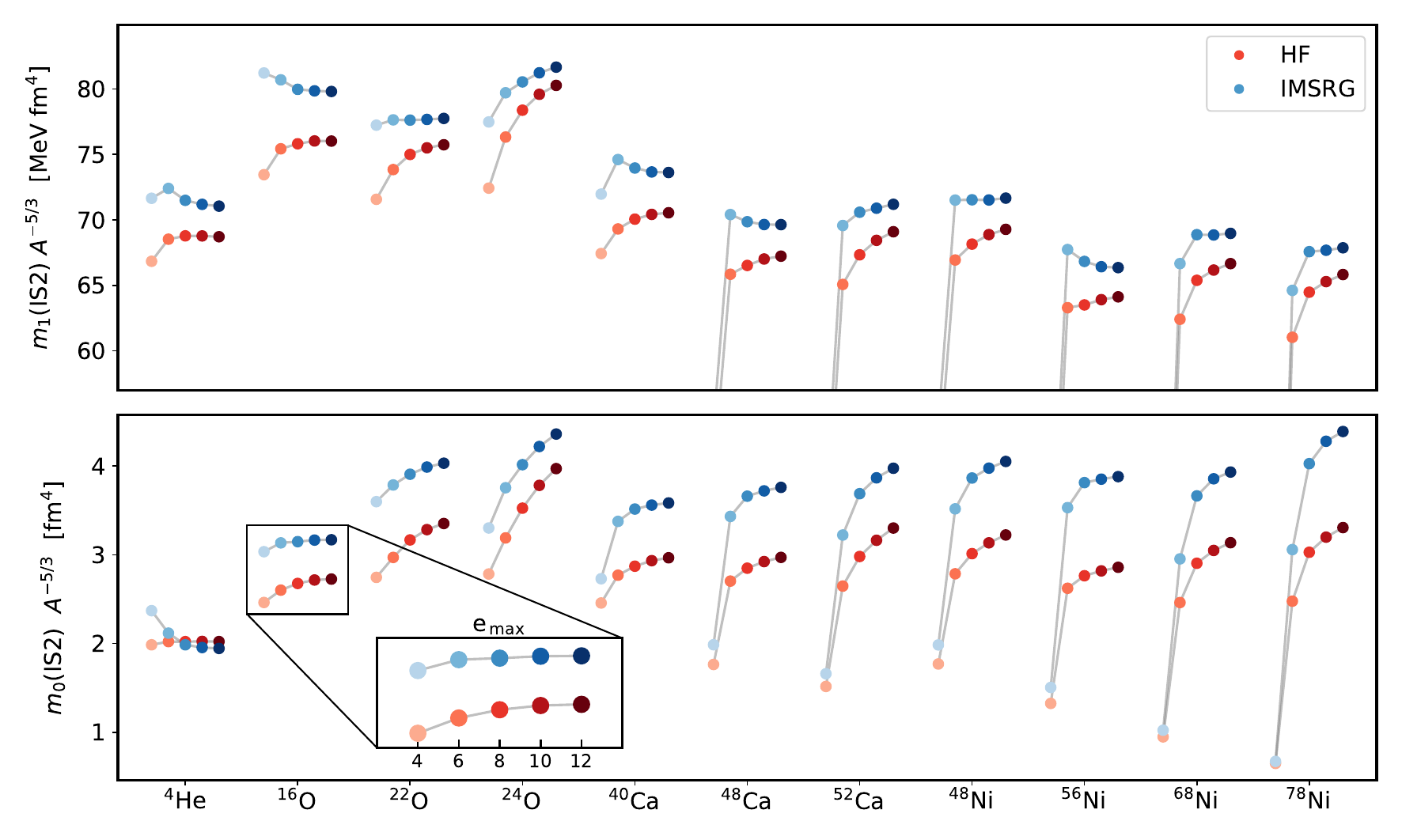}
    \caption{Same as Fig.~\ref{fig:L1_emax_conv} but for the isoscalar quadrupole (IS2) response.}
    \label{fig:L2_emax_conv}
\end{figure*}

\section{Model-space convergence}
\label{app:ms_conv}

\begin{figure*}
    \centering
    \includegraphics[width=\textwidth]{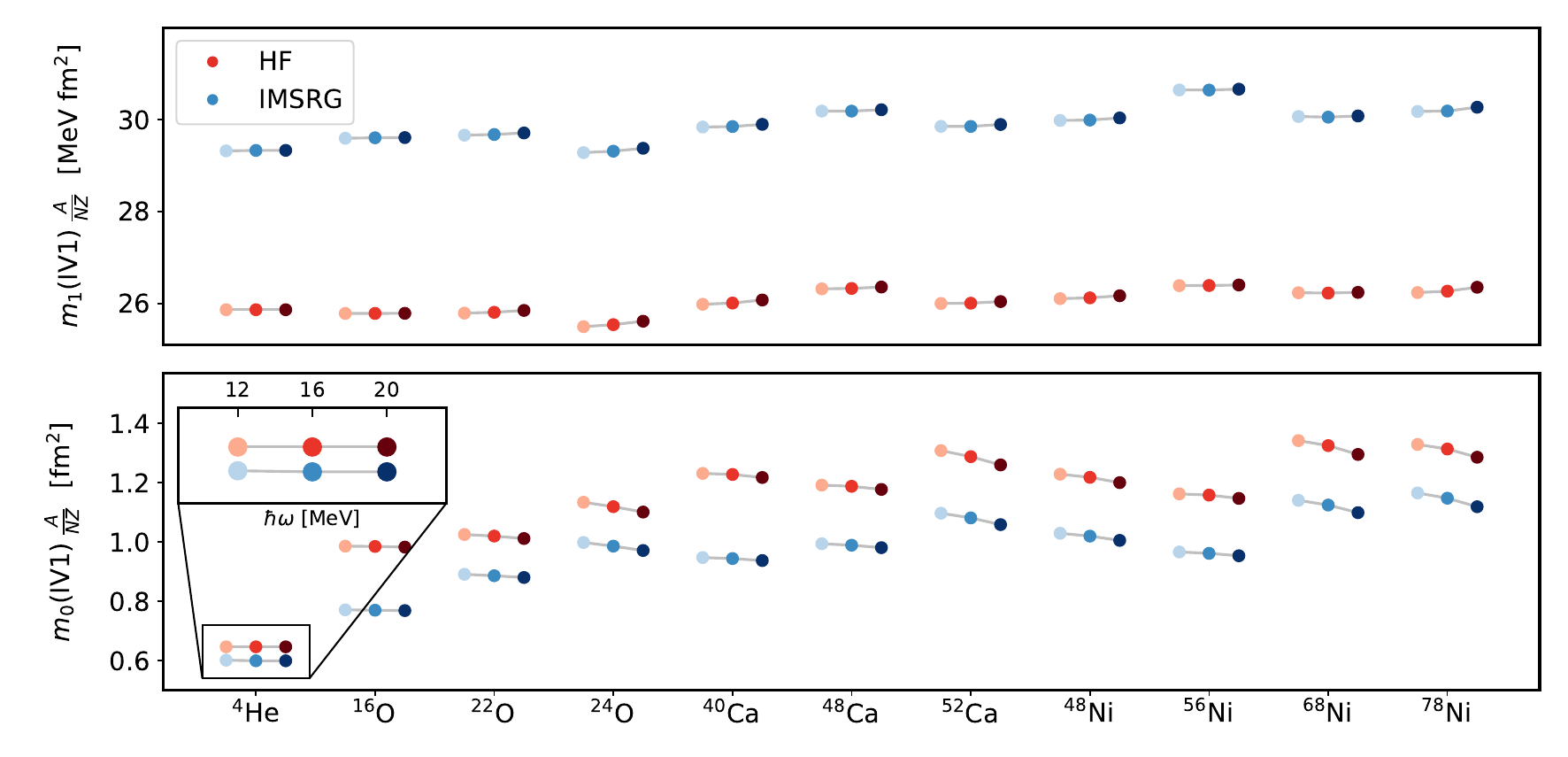}
    \caption{Dependence of $m_1$ (top panel) and $m_0$ (bottom panel) for the isovector dipole (IV1) response on the HO frequency $\hbar\omega$ at the HF and IMSRG(2) level. For each isotope, the results with increasing $\hbar\omega$ are shown, as highlighted in the inset. The values of $m_1$ and $m_0$ are rescaled by a factor $\frac{A}{NZ}$ in order to show all nuclei on the same scale [see Eq.~\eqref{eq:trk}]. Results are given for the 1.8/2.0 (EM) interaction using \Xm{e}=12.}
    \label{fig:L1_hw_conv}
\end{figure*}

Physical observables should not depend on the specific basis used in numerical implementations. However, due to the finite size of the Hilbert space in actual implementations, a residual dependence is to be expected and quantified. The same analysis performed in Sec.~\ref{sec:ms_conv} for the model space convergence of the isovector dipole response is repeated here for the isoscalar monopole (Fig.~\ref{fig:L0_emax_conv}) and quadrupole (Fig.~\ref{fig:L2_emax_conv}) response. We report the largest relative error for the studied cases. For the IS0 response, the largest relative difference for $m_1$ is 1.4\% at the HF level for $^{52}$Ca and 1.0\% at the IMSRG(2) level for $^4$He. The spread is larger when studying $m_0$, with a difference in $^{24}$O of 6.8\% (HF) and 5.2\% (IMSRG(2)). The IS2 response displays a relative error always smaller than 1\% for $m_1$ for the range of nuclei studied, while the largest error for $m_0$ is 4.8\% and 3.2\% in $^{24}$O, respectively at the HF and IMSRG(2) level. The dependence of the IV1 response on the HO parameter $\hbar\omega$ has also been investigated. The analysis is displayed in Fig.~\ref{fig:L1_hw_conv}, and we refer to the discussion in the main in Sec.~\ref{sec:ms_conv}.

\bibliography{biblio}

\end{document}